\documentclass[aps,prd,reprint,twocolumn,floatfix,preprintnumbers,nofootinbib]{revtex4-2}
\usepackage{graphicx,color,xcolor}
\usepackage{amsmath,amssymb,amsfonts,amsthm}
\usepackage{wasysym}
\usepackage[caption=false]{subfig}
\usepackage[colorlinks=True, citecolor=blue, urlcolor=blue, linkcolor=blue]{hyperref}
\usepackage{ulem}
\usepackage{comment}
\newcommand{\be}{\begin{equation}}
\newcommand{\ee}{\end{equation}}
\newcommand{\ba}{\begin{eqnarray}}
\newcommand{\ea}{\end{eqnarray}}
\newcommand{\bs}{\boldsymbol}
\newcommand{\nn}{\nonumber}

\newcommand{\jmax}{j_\mathrm{max}}

\begin{document}
\title{Entanglement Entropy of ($\mathbf{2+1}$)-Dimensional SU(2) Lattice Gauge Theory on Plaquette Chains}
\author{Lukas Ebner}
\affiliation{Institut f\"ur Theoretische Physik, Universit\"at Regensburg, D-93040 Regensburg, Germany}
\author{Berndt M\"uller}
\affiliation{Department of Physics, Duke University, Durham, North Carolina 27708, USA}
\author{Andreas Sch\"afer}
\affiliation{Institut f\"ur Theoretische Physik, Universit\"at Regensburg, D-93040 Regensburg, Germany}
\author{Clemens Seidl}
\affiliation{Institut f\"ur Theoretische Physik, Universit\"at Regensburg, D-93040 Regensburg, Germany}
\author{Xiaojun Yao}
\affiliation{InQubator for Quantum Simulation, Department of Physics,
University of Washington, Seattle, Washington 98195, USA}
\date{\today}
\preprint{IQuS@UW-21-070}
\begin{abstract}
We study the entanglement entropy of Hamiltonian SU(2) lattice gauge theory in $2+1$ dimensions on linear plaquette chains and show that the entanglement entropies of both ground and excited states follow Page curves. The transition of the subsystem size dependence of the entanglement entropy from the area law for the ground state to the volume law for highly excited states is found to be described by a universal crossover function. Quantum many-body scars in the middle of the spectrum, which are present in the electric flux truncated Hilbert space, where the gauge theory can be mapped onto an Ising model, disappear when higher electric field representations are included in the Hilbert space basis. This suggests the continuum $(2+1)$-dimensional SU(2) gauge theory does not have such scarred states.
\end{abstract}
\maketitle
{\it Introduction.} Entanglement is a peculiar property of quantum systems that has no classical analog~\cite{Eisert:2008ur,Nishioka:2018khk}. Entanglement entropy of a subsystem in a pure state quantifies the amount of entanglement between the subsystem and its complement. Calculations of entanglement entropy have wide physical applications such as understanding the Bekenstein-Hawking radiation from a quantum perspective~\cite{Bombelli:1986rw,Almheiri:2019psf,Penington:2019npb,Almheiri:2019hni}, characterizing phases of many-body systems~\cite{Kitaev:2005dm,Levin:2006zz,Li:2008kda}, understanding thermalization of isolated quantum systems~\cite{Majidy:2022kzx,Mueller:2021gxd}, detecting quantum many-body scars~\cite{Turner:2018kjz,Ho:2018rum,Moudgalya:2018nuz,Iadecola:2019jdz,Banerjee:2020tgz,Aramthottil:2022jvs,Evrard:2023qnu} and proving monotonic renormalization group flow for certain physical quantities~\cite{Casini:2004bw,Casini:2012ei,Itou:2015cyu,Rabenstein:2018bri}.

For quantum field theories, studies of entanglement entropy have been largely limited to $(1+1)$-dimensional theories, such as conformal field theories, and have generally made use of the replica method~\cite{Callan:1994py,Calabrese:2004eu} or holographic techniques in cases where the field theory has a gravity dual~\cite{Ryu:2006bv,Klebanov:2007ws}. For gauge theories, these techniques are not easily applied. One difficulty, which is manifest on a lattice, is the appearance of gauge non-invariant states on the edge of the subsystem (or its complement), i.e., on the entangling surface~\cite{Buividovich:2008gq,Donnelly:2011hn}. Many studies have noted this problem and described methods of adding extra states in order to factorize the complete Hilbert space into two sectors, one for the subsystem and one for its complement~\cite{Casini:2013rba,Ghosh:2015iwa,Aoki:2015bsa,Soni:2015yga,Agarwal:2016cir}. A numerical study using the replica trick in the Euclidean path integral can be found in~\cite{Buividovich:2008kq}, and a calculation using tensor networks in~\cite{Cataldi:2023xki}.

Here we provide a construction applicable to the Hamiltonian formulation of lattice gauge theory~\cite{PhysRevD.11.395}. The advantage of the Hamiltonian formulation is that we can study excited pure quantum states and not only thermal ensembles. This will allow us, e.g., to obtain the Page curve for excited states. We consider the $(2+1)$-dimensional SU(2) gauge theory on linear plaquette chains as an example. The division into a subsystem and its complement in our construction is carried out in the electric basis and cuts through links. The resulting reduced density matrix is invariant under time-independent gauge transformations at each vertex. As consistency checks, we will show analytically that the entanglement entropy of a subsystem is equal to that of its complement and provide numerical evidence that our method gives symmetric Page curves for both ground and excited pure quantum states. 

Our construction also allows us to search for quantum many-body scar states among all the eigenstates by studying their entanglement entropies.
Such states are identified as those highly excited eigenstates with low entanglement entropies that weakly break ergodicity and violate the strong eigenstate thermalization hypothesis. Scarred states delay or inhibit the thermalization of a many-body system. 

So the current work complements our previous studies that investigate the eigenstate thermalization hypothesis~\cite{Yao:2023pht,Ebner:2023ixq}. Major results of our current work are as follows:
\begin{itemize}
\item We define partitions of a lattice system in a way compatible with the local Gauss law;

\item We calculate entanglement entropies of pure ground and highly excited states and confirm the transition from the area law to the volume law smoothly follows a universal crossover function;

\item We find quantum many-body scars when the local Hilbert space is truncated at $j_{\rm max}=\frac{1}{2}$, which disappear with higher $j_{\rm max}$ truncation.
\end{itemize}

{\it Hamiltonian.}
For simplicity, we consider the $(2+1)$-dimensional SU(2) lattice gauge theory on a square plaquette chain. (It is straightforward to extend our study to the spatial plane by using a honeycomb lattice~\cite{Muller:2023nnk}, however, this would require significantly larger computing resources.) The Kogut-Susskind Hamiltonian of the lattice gauge theory is~\cite{PhysRevD.11.395}
\begin{align}
\label{eqn:H}
H = \frac{g^2}{2}\sum_{\rm L} (E_i^a)^2 - \frac{2}{a^2g^2} \sum_{\rm P} {\rm Tr}\left[ \prod_{({\bs n},\hat{i})\in {\rm P}} 
U({\bs n},\hat{i})\right] \,,
\end{align}
where $\rm L$ denotes links, $\rm P$ plaquettes, and the product in the second term is over the four oriented links of a plaquette. $\bs n$ and $\hat{i}$ indicate the location and direction of a link. $a$ denotes the lattice spacing and $g$ is the gauge coupling. Physical states are those that satisfy Gauss law at each vertex and can be fully expressed in the electric basis $|j,m_L,m_R\rangle$ that describes states on each link~\cite{Byrnes:2005qx,Zohar:2014qma,Liu:2021tef}, where $m_L,m_R$ are the third components of the angular momentum quantum number $j$ on the two ends of a link. In this basis, the electric energy $(E_i^a)^2$ is diagonal with eigenvalues $j(j+1)$. The $m_L,m_R$ dependence can be integrated out by imposing Gauss law at each vertex. As a result, the matrix elements of the plaquette term (which gives the shifted magnetic energy) can be expressed in terms of Wigner $6$-$j$ symbols~\cite{Klco:2019evd}. With $j$ truncated at $\jmax$ on each link, the Hilbert space on a given lattice is finite dimensional, and the Hamiltonian can be exactly diagonalized numerically.

Much longer plaquette chains can be accommodated if only states with $j\le j_{\rm max}=\frac{1}{2}$ are taken into account. In this constrained basis the SU(2) theory on a plaquette chain becomes equivalent to an Ising model with the Hamiltonian~\cite{Yao:2023pht} (see also~\cite{Hayata:2021kcp,ARahman:2022tkr})
\begin{align}
H = \sum_{i=0}^{N-1} \left( J \sigma_i^z \sigma_{i+1}^z - 2J \sigma_i^z + h_x \nu_i \sigma_i^x \right) \, ,
\label{eq:H_Ising}
\end{align} 
where $J = -3g^2/16$, $h_x = (ag)^{-2}$, $\nu_i = (i/\sqrt{2})^{\sigma_{i-1}^z+\sigma_{i+1}^z}$. We impose periodic boundary conditions $\sigma_N^z=\sigma_0^z$. In the following, we study both the Ising limit and Hilbert spaces with $j_{\rm max}>\frac{1}{2}$. 
In all cases our calculations involve Hilbert spaces of dimension $O(10^4)$.

\begin{figure}[t]  
\centering
\includegraphics[width=0.5\textwidth]{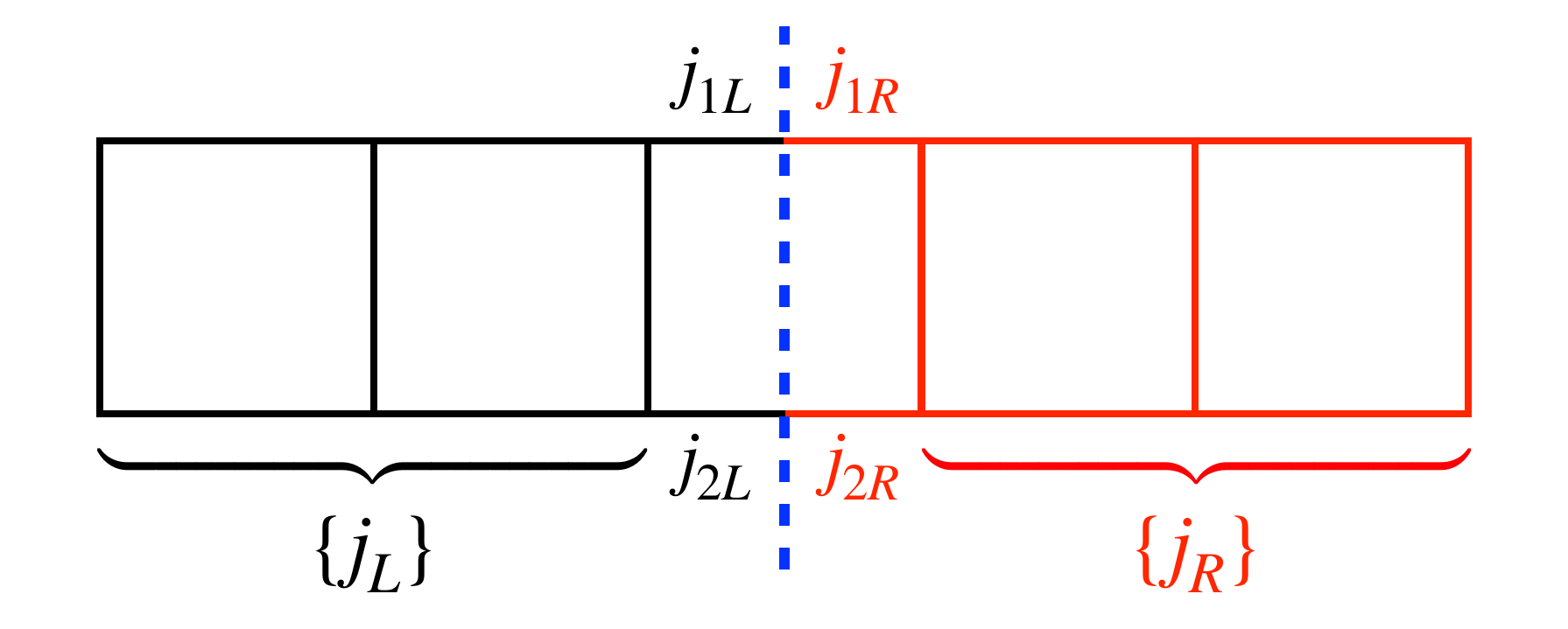}
\caption{A five-plaquette chain cut into left (black) and right (red) parts through two horizontal links. The state on the left is labeled by the collection of $j$ values $\{j_L\}$ on the four horizontal and three vertical unbroken links and the $j$ values $j_{1L}$, $j_{2L}$ for the two dangling links. Similarly for the state on the right. $j_{1L}=j_{1R}$ and $j_{2L}=j_{2R}$ since they form the same link before the cut.}
\label{fig:subsystem}
\end{figure}

{\it Entanglement entropy.} 
The entanglement entropy $S_A$ associated with a segment $A$ of the entire lattice supporting a pure quantum state $|\psi\rangle$ is given by 
\begin{equation}
S_A = -{\rm Tr}\,(\rho_A\ln\rho_A) \,,
\end{equation}
where $\rho_A={\rm Tr}_{A^c} |\psi\rangle\langle\psi|$ is the reduced density matrix of the state on the segment $A$, and $A^c$ denotes the complement of $A$. Consider, e.g., a state on a five-plaquette chain as shown in Fig.~\ref{fig:subsystem}. Here we express the state in the electric basis as $|\{ j\} \rangle$, where the set $\{j\}$ denotes the collection of eigenvalues $j$ of the electric field on each link. For physical states, the three $j$ values at each vertex must form a SU(2) singlet. We divide the chain into left and right segments by cutting through two horizontal links and decompose an arbitrary state $|\psi\rangle$ in terms of a state living on the left segment and another living on the right segment:
\begin{align}
|\psi\rangle =\ & c_{\{j_L\}j_{1L}j_{2L}\{j_R\}j_{1R}j_{2R}} \delta_{j_{1L}j_{1R}} \delta_{j_{2L}j_{2R}} \nn\\
& |\psi_{\{j_L\}j_{1L}j_{2L}}\rangle \otimes |\psi_{\{j_R\}j_{1R}j_{2R}}\rangle \,,
\end{align}
where $c$ denotes the combinatorial coefficients and repeated indexes are summed. The delta functions ensure that the two dangling links on the top (bottom) at the boundary have the same $j$ value since they form the same link before the cut. The reduced density matrix of the left chain can be obtained by tracing out $\{j_R\},j_{1R},j_{2R}$
\begin{align}
\!\!\! \rho_L =\ &  c_{\{j_L\}j_{1L}j_{2L}\{j_R\}j_{1L}j_{2L}} c_{\{j_L\}'j_{1L}j_{2L}\{j_R\}j_{1L}j_{2L}} \nn\\
&  |\psi_{\{j_L\}j_{1L}j_{2L}}\rangle \langle \psi_{\{j_L\}'j_{1L}j_{2L}}| \,.
\end{align}
The reduced density matrix is a direct sum
\begin{align}
\rho_L = \bigoplus_{j_{1L},\,j_{2L}} \rho_L(j_{1L}, \,j_{2L}) \,,
\end{align}
where the trace of $\rho_L(j_{1L}, \,j_{2L})$ is not necessarily unity.
A similar expression can be written down for $\rho_R$. Since Gauss law has been integrated out in the $\{j\}$ basis at each vertex and the cutting does not involve any vertex, the reduced density matrix is still invariant under time-independent gauge transformations at each vertex.

\begin{figure*}[t]  
\centering
\subfloat[$j_{\rm max}=1.5, N=6, g^2a=1.2$.\label{fig:EE_N5_ag2_1.2_vs_Nsub}]{%
  \includegraphics[width=0.33\linewidth]{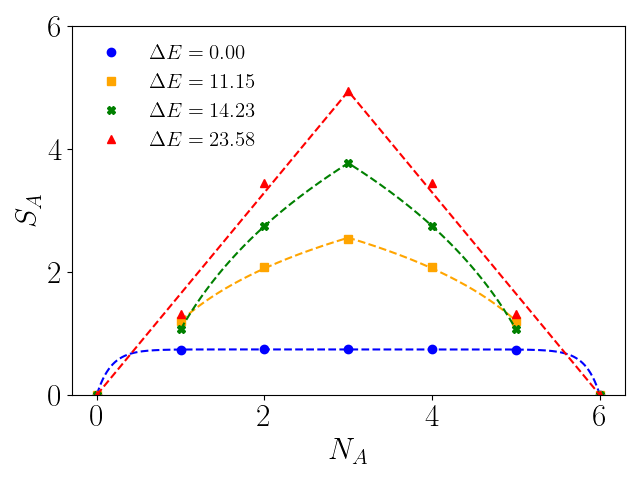}%
}\hfill
\subfloat[$j_{\rm max}=1, N=8, g^2a=0.8$.\label{fig:EE_N7_ag2_0.8_vs_Nsub}]{%
  \includegraphics[width=0.33\linewidth]{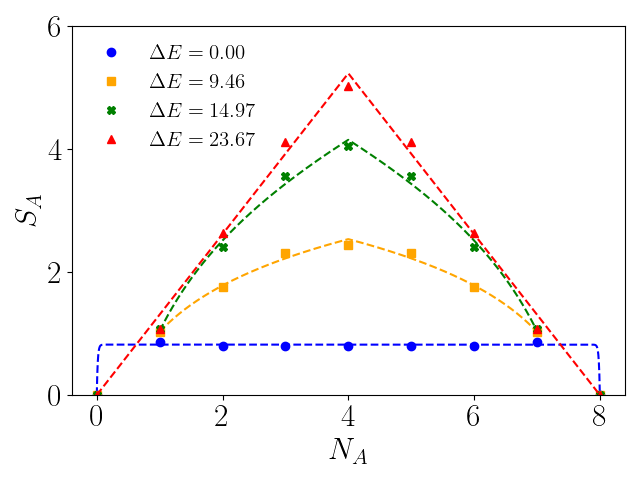}%
}\hfill
\subfloat[$j_{\rm max}=\frac{1}{2}, N=17, k=0, g^2a=1.2$.\label{fig:EE_N17_k0_ag2_1.2_vs_Nsub}]{%
  \includegraphics[width=0.33\linewidth]{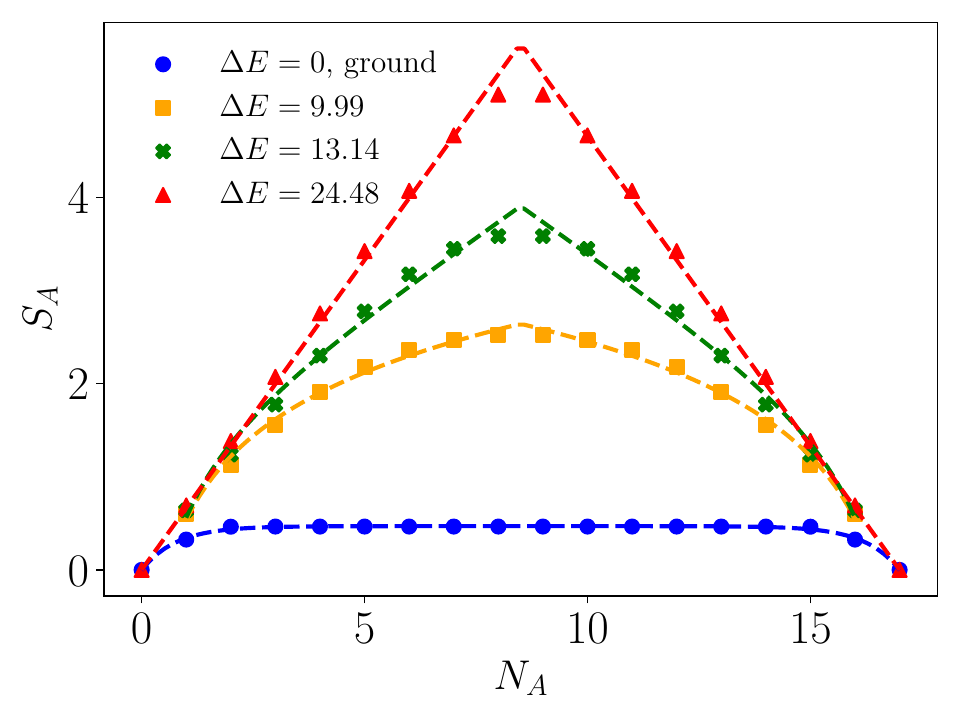}%
}
\caption{Entanglement entropy as a function of the subsystem size $N_A$ on (a) an aperiodic five-plaquette chain with $j_{\rm max}=1.5$, (b) an aperiodic seven-plaquette chain with $j_{\rm max}=1$, and (c) a periodic 17-plaquette chain in the zero momentum sector with $j_{\rm max}=\frac{1}{2}$. The states are labeled by their energies relative to the ground state energy $\Delta E=E-E_0$ in lattice units.}
\label{fig:page_N_5_7_17}
\end{figure*}

The second-order R\'enyi entropy is given by
\begin{align}
\!\!\!\!{\rm Tr}(\rho_L^2) = \ 
& c_{\{j_L\}j_{1L}j_{2L}\{j_R\}j_{1L}j_{2L}} c_{\{j_L\}'j_{1L}j_{2L}\{j_R\}j_{1L}j_{2L}} \nn\\
& c_{\{j_L\}'j_{1L}j_{2L}\{j_R\}'j_{1L}j_{2L}} c_{\{j_L\}j_{1L}j_{2L}\{j_R\}'j_{1L}j_{2L}} \,.
\end{align}
Since $j_{1L}=j_{1R}$ and $j_{2L}=j_{2R}$, by replacing the sums over $j_{1L},j_{2L}$ with $j_{1R},j_{2R}$ we obtain
\begin{align}
{\rm Tr}(\rho_L^2) = {\rm Tr}(\rho_R^2) \,.
\end{align}
Similarly one can prove
\begin{align}
\label{eqn:L=R}
{\rm Tr}(\rho_L^n) = {\rm Tr}(\rho_R^n) \,,
\end{align}
for any integer $n>2$. When the Hilbert space size is finite, as is the case on a finite lattice with a finite $j_{\rm max}$, Eq.~\eqref{eqn:L=R} implies that $\rho_L$ and $\rho_R$ have the same eigenvalues and thus have the same entanglement entropy $S_L=S_R$. We will later confirm this result numerically. The calculation of $S_A$ can be simplified when $j_{\rm max}=\frac{1}{2}$, as explained in Appendix~\ref{app:a}.

{\it Results.} We first study the entanglement entropy of the ground state and various excited states as a function of the subsystem size for three lattice configurations: (a) a five-plaquette chain with $j_{\rm max}=1.5$, $g^2a=0.8$ and closed boundary conditions $j_{\rm ext}=0$ on the external links; (b) a seven-plaquette chain with the same parameters as (a) except for $j_{\rm max}=1$; and (c) a periodic 17-plaquette chain in the zero momentum ($k=0$) sector with $j_{\rm max}=\frac{1}{2}$ and $g^2a=1.2$. 
We note that on an aperiodic chain the number of vertices along the top link is one more than that of plaquettes and thus the subsystem length may take the values $N_A\in\{0,...,N\}$, where $N$ is the number of top vertices. Our results are shown in Fig.~\ref{fig:page_N_5_7_17}, where all the numerical points are calculated without using the symmetry between $N_A$ and $N-N_A$, demonstrating the expected Page curves.

\begin{figure}[t]  
\centering
\includegraphics[width=0.45\textwidth]{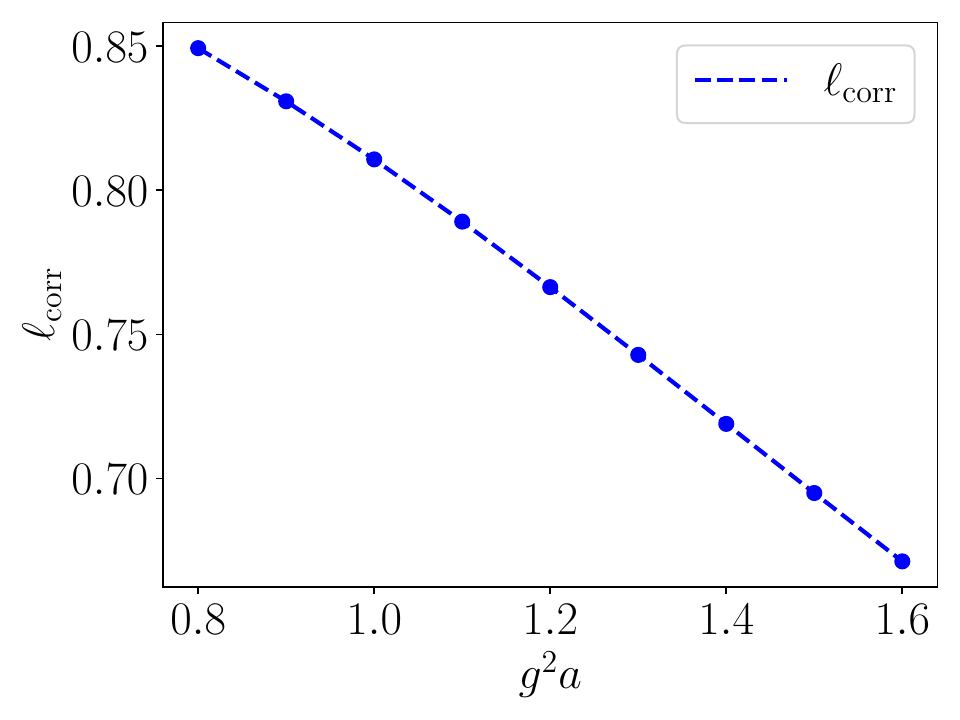}
\caption{Correlation length as a function of the coupling extracted from the ground state entanglement entropy in case (c).}
\label{fig:lcorr}
\end{figure}

Consistent with the expectation for gapped $(1+1)$-dimensional systems~\cite{Hastings:2007iok} our results for the ground state satisfy the area law, while those for the highly excited states follow the volume law. To demonstrate this for case (c), we fit the ground-state result by the function
\begin{align}
S_{\rm area}(N_A) = b_0-b_1(e^{-N_A/\ell_{\rm corr}} + e^{-(N-N_A)/\ell_{\rm corr}}) \,,
\end{align}
where $b_0$ and $b_1$ are constants and $\ell_{\rm corr}$ is the correlation length of the system. We use the exponential dependence on $N_A$ since the spectrum of our Hamiltonian is gapped and thus not conformal.\footnote{Of course, the discrete lattice would break a possibly existing conformal symmetry, but it is known from numerical studies that the continuum limit of the SU(2) gauge theory exhibits a mass gap (the glueball mass).} Fits with polynomial or logarithmic functions do not generate good results. The fitted values are $b_0=0.470$, $b_1=0.472$ and $\ell_{\rm corr}=0.766$. We also study the coupling dependence of the correlation length, which is shown in Fig.~\ref{fig:lcorr} and exhibits an approximately linear dependence. We note that by requiring the physical correlation length $a\ell_{\rm corr}$ to be invariant as $a$ changes, we can obtain a renormalization group equation for $g^2a$. Our plaquette chains for $j_{\rm max}>\frac{1}{2}$ are currently too short to permit a reliable determination of $\ell_{\rm corr}$.

\begin{figure}[t]  
\centering
{%
  \includegraphics[width=0.45\textwidth]{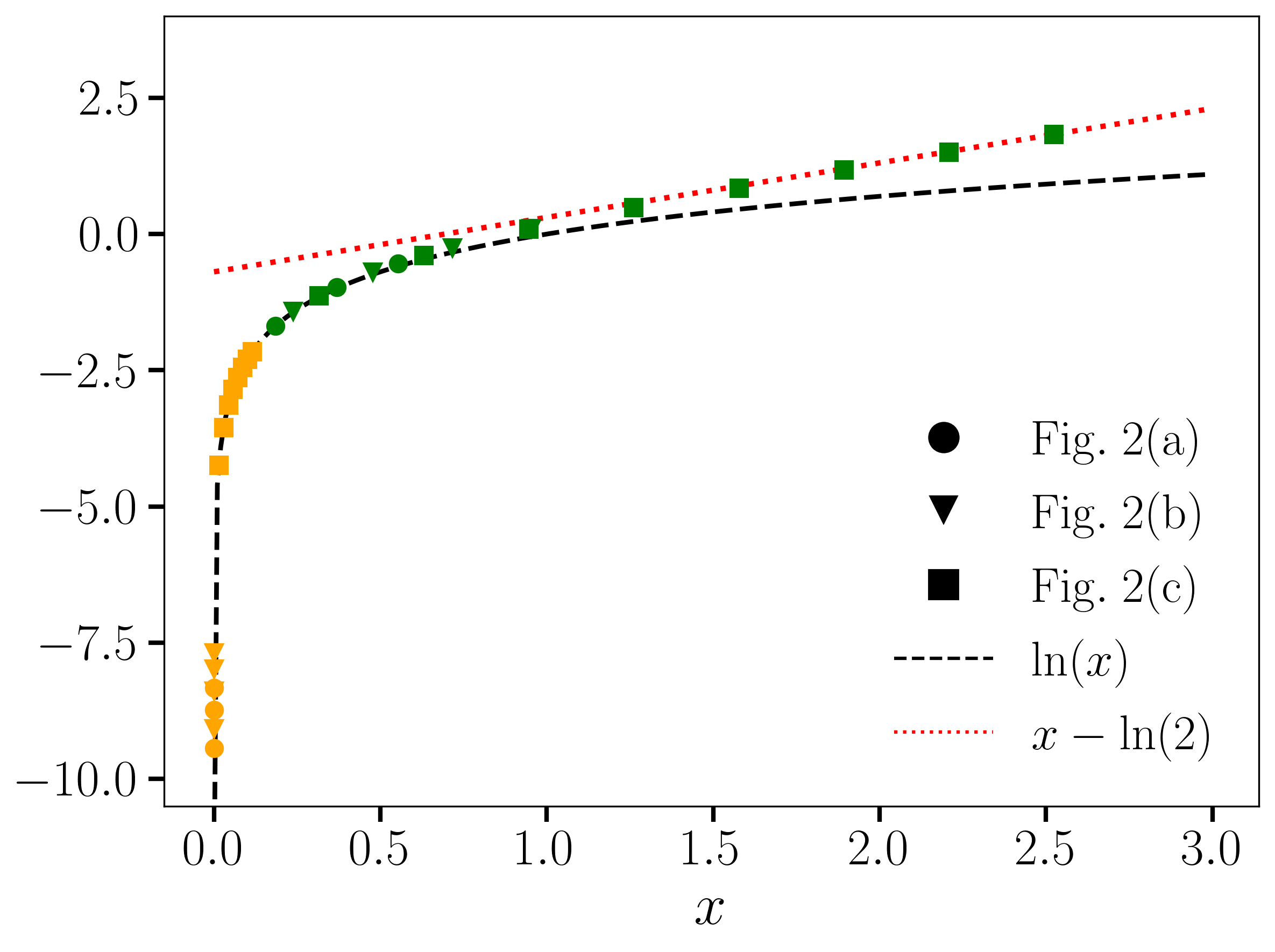}%
}

\caption{The function $\ln[\sinh(x)]$ evaluated at the points $x=c_1^{-1}N_A$, where $c_1$ is taken from the fit results to the orange and green states in Figs.~\ref{fig:EE_N5_ag2_1.2_vs_Nsub}, ~\ref{fig:EE_N7_ag2_0.8_vs_Nsub}, ~\ref{fig:EE_N17_k0_ag2_1.2_vs_Nsub} here depicted by the orange and green markers. The subsystem sizes are $N_A=1,..,\lfloor \frac{N}{2} \rfloor$. Additionally the asymptotic behavior of $\ln[\sinh(x)]$ for small (black dashed) and large (red dotted) $x$ is shown.}
\label{fig:transition}
\end{figure}

\begin{figure*}[t]  
\centering
\subfloat[$N=17$.\label{fig:EE_density_N17_k0_Nsub8_ag2_1.6}]{%
  \includegraphics[width=0.33\linewidth]{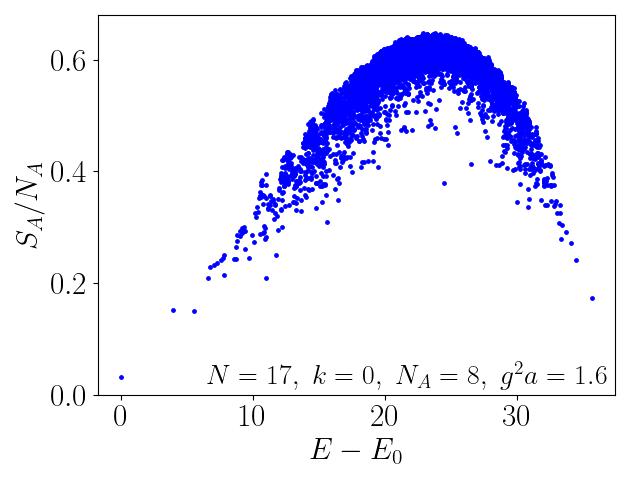}%
}\hfill
\subfloat[$N=19$.\label{fig:EE_density_N19_k0_Nsub9_ag2_1.6}]{%
  \includegraphics[width=0.33\linewidth]{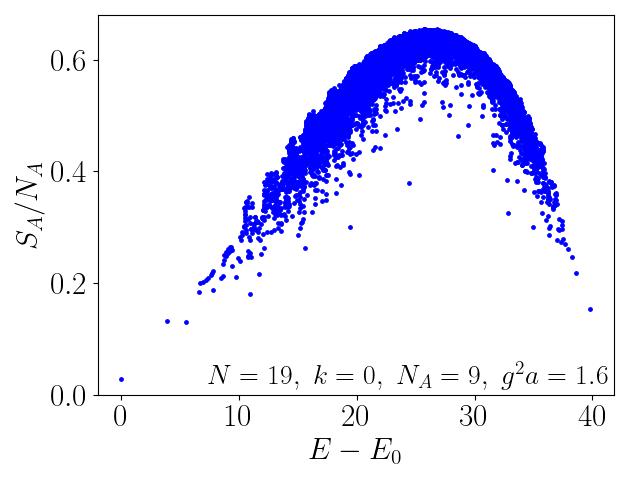}%
}\hfill
\subfloat[$N=17$.\label{fig:EE_N17_k0_ag2_1.6_QMBS}]{%
  \includegraphics[width=0.33\linewidth]{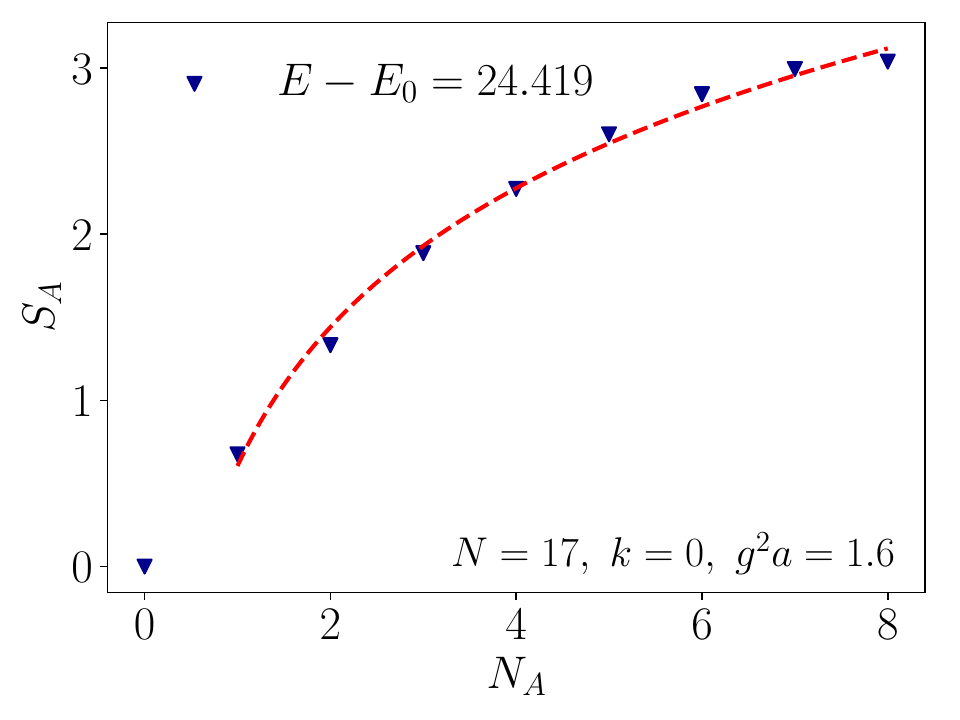}%
}
\caption{Half-chain entanglement entropies of all energy eigenstates in the $k=0$ sector on periodic (a) $N=17$ and (b) $N=19$ plaquette chains with $j_{\rm max}=\frac{1}{2}, g^2a=1.6$. (c) Entanglement entropy of the outsider state in the $N=17,k=0$ case whose energy is 24.419 above the ground state as a function of the subsystem size. Red dashed line is a fit of Eq.~\eqref{eq:Scross} with $(c_0,c,c_1)=(0.61, 3.61, 39.45)$. The large value of $c_1$ indicates the $N_A$ dependence is logarithmic.}
\label{fig:EE_N17_N19_jmax0.5_QMBS}
\end{figure*}

We fit the results for the most highly excited state considered in all three cases, which corresponds to the peak of the level density (i.e.\ infinite temperature) and is shown by red triangles in Fig.~\ref{fig:page_N_5_7_17} with a volume law function
\begin{align}
S_{\rm vol}(N_A) = s N_A\theta\Big(\frac{N}{2}-N_A\Big) + s(N-N_A)\theta\Big(N_A-\frac{N}{2}\Big)\,,
\end{align}
where $s$ is a constant that can be interpreted as the entropy density. The fitted value in case (c) is $s=0.665$, which is very close to the entropy constraint $S/N\leq\ln2$. For case (a) and (b), the fitted values are $s=1.646$ and $s=1.307$, respectively.

The transition from the area law in the ground state to the volume law in the highly excited state can be represented by a crossover function that was derived for $(1+1)$-dimensional conformal field theory in Ref.~\cite{PhysRevLett.127.040603}
\begin{align}
\label{eq:Scross}
&S_{\rm cross}(N_A) = c_0 + \frac{c}{3}\ln[c_1\sinh(c_1^{-1}N_A)] \theta\Big(\frac{N}{2}-N_A\Big) \nn\\
& \qquad + \frac{c}{3}\ln\{c_1\sinh[c_1^{-1} (N-N_A)]\} \theta\Big(N_A-\frac{N}{2} \Big) \nn\\
& \qquad = c_0+\frac{c}{3}\ln(c_1) + \frac{c}{3} \Big\{ \ln[\sinh(c_1^{-1}N_A)] \theta\Big(\frac{N}{2}-N_A\Big) \nn\\
& \qquad + \ln\{\sinh[c_1^{-1} (N-N_A)]\}\theta\Big(N_A-\frac{N}{2} \Big)\Big\} \,,
\end{align}
where $c_0$ is a constant that cancels the divergences at $N_A=0$ and $N_A=N$.\footnote{In our following fit study, we only use the data points at $N_A\in[1,N-1]$ due to the divergence in the $\ln(\sinh)$ function. So the fitted result of $c_0$ is always finite. It will become infinite in the limit $a\to0$.} $c_1$ and $c$ are other fit parameters. $c$ has the physical meaning of the central charge for conformal field theories. Our use of this function is motivated by the fact that at high temperature the continuum SU(2) gauge theory is known to be approximately conformal. Furthermore, Figs.~\ref{fig:EE_N17_k0_ag2_1.2_vs_Nsub} and~\ref{fig:lcorr} show that the energies of the excited states studied are much larger than the inverse of the correlation length $1/\ell_{\rm corr}$ for case (c). We conclude that the excited states studied here truly correspond to the high-temperature regime. Excluding the singular points $N_A=0$ and $N_A=N$, our fitted results are shown in Table~\ref{tab:Table1}.
\begin{table}[htb]
\centering
\begin{tabular}{|c|c|c|}
\hline
  & Orange & Green \\
\hline
Fig.~\ref{fig:EE_N5_ag2_1.2_vs_Nsub} & (1.20, 3.70, 12476.0) & (1.05, 7.09, 5.42) \\
Fig.~\ref{fig:EE_N7_ag2_0.8_vs_Nsub} & (1.03, 3.26, 8772.0) & (1.02, 6.12, 4.19) \\
Fig.~\ref{fig:EE_N17_k0_ag2_1.2_vs_Nsub} & (0.54, 2.95, 69.5) & (0.54, 3.22, 3.17) \\
\hline
\end{tabular}
\caption{Fitted parameter values $(c_0,c,c_1)$ using Eq.~(\ref{eq:Scross}) for the subsystem entanglement entropies of the excited states marked as orange and green in Fig.~\ref{fig:page_N_5_7_17}.}
\label{tab:Table1}
\end{table}

In order to illustrate the transition to the volume law in the three panels of Fig.~\ref{fig:page_N_5_7_17}, we expose their functional dependence on $N_A$. To do so, we subtract the constant terms in Eq.~\eqref{eq:Scross}, divide the results by $\frac{c}{3}$, and restrict to the region $N<N_A/2$. The remaining part of Eq.~\eqref{eq:Scross} has the form $\ln[\sinh(x)]$ with $x=c_1^{-1}N_A$. The numerical results for the orange and green fits in Fig.~\ref{fig:page_N_5_7_17} are plotted as points in Fig.~\ref{fig:transition} as a function of $x$. We also show the asymptotic forms $\ln(x)$ for the logarithm enhanced area law (black dashed curve) and $x-\ln(2)$ for the volume law (red dotted line). The approach of the numerical points to the linear behavior marks the transition to the volume law, which is seen to occur around $x\approx 1$.

{\it Scars.} Next we study the half-chain ($N_A=\lfloor \frac{N}{2}\rfloor$) entanglement entropy for each eigenstate and search for quantum many-body scar states. The results on a periodic plaquette chain with $j_{\rm max}=\frac{1}{2}$ and $g^2a=1.6$ are shown in Fig.~\ref{fig:EE_N17_N19_jmax0.5_QMBS}. At this coupling the system is non-integrable and exhibits an equal partition between the electric and magnetic energies (see Appendix~\ref{app:b}) which is a property of the continuum SU(2) gauge theory at thermal equilibrium. An eigenstate in the middle of the spectrum with excitation energy $E-E_0 = 24.419$ above the ground state is found to be a low-entanglement state whose entanglement entropy deviates strongly from all its neighbors. This state is identified as a quantum many-body scar due to two observations. First, we find no significant volume dependence of the deviation when comparing the results obtained from the $N=17$ and $N=19$ chains and Furthermore, the entanglement entropy of this state exhibits a logarithmic dependence on the subsystem size, which resembles a lower-energy eigenstate and differs greatly from its neighbors that follow the volume law, as exemplified by the red curve in Fig.~\ref{fig:EE_N17_k0_ag2_1.2_vs_Nsub}. Figure~\ref{fig:EE_density_N19_k0_Nsub9_ag2_1.6} also suggests the existence of a tower of quantum many-body scars below $E-E_0=24.42$.
\begin{figure*}[t]  
\centering
\subfloat[Scarred eigenstate.\label{fig:QMBS_9950_N19_k0_ag2_1.6}]{%
  \includegraphics[width=0.45\textwidth]{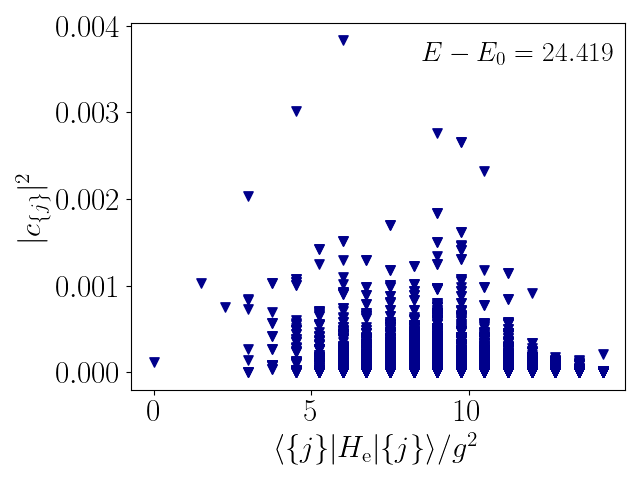}%
}
\hfill
\subfloat[Typical eigenstate.\label{fig:QMBS_9945_N19_k0_ag2_1.6}]{%
  \includegraphics[width=0.45\textwidth]{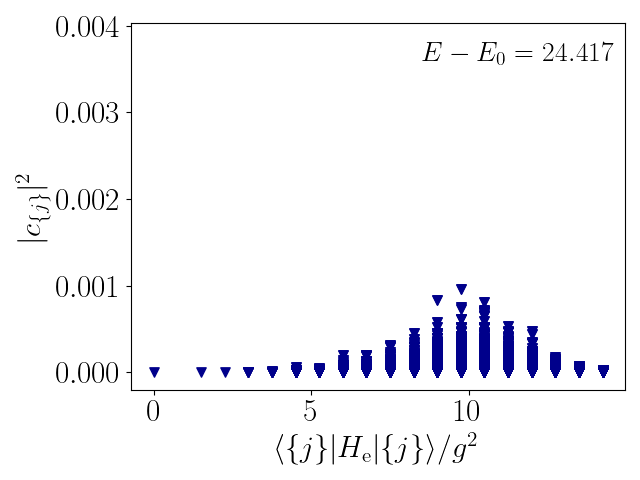}%
}
\caption{Wavefunction components of the scarred eigenstate and a typical eigenstate around the energy $E-E_0=24.42$ in the electric basis in the $k=0$ sector on a periodic $N=19$ plaquette chain with $j_{\rm max}=\frac{1}{2}$ and $g^2a=1.6$. Their half-chain entanglement entropies are $3.42$ and $5.76$, respectively. The $x$-axis labels the electric energy of the basis state which is zero when all $j$ values are $0$. We fix the $y$-axis scale to be the same in both plots for visual comparison.}
\label{fig:scar_state}
\end{figure*}

To understand the nature of the scarred eigenstate, we study its wavefunction components in the electric basis $|\psi\rangle = c_{\{j\}} |{\{j\}}\rangle$ where ${\{j\}}$ is a collection of $j$ values on all the links and implicitly summed over. We plot $|c_{\{j\}}|^2$ and the associated electric energy of the basis state in Fig.~\ref{fig:scar_state} for the scarred eigenstate and a typical state around the energy $E-E_0=24.42$ for the $k=0$ sector of the periodic $N=19$ plaquette chain with $j_{\rm max}=\frac{1}{2}$. Weight factors $|c_{\{j\}}|^2$ in typical eigenstates are peaked around a given electric energy but are not dominated by some few single states. On the other hand, the scarred eigenstate has larger components in some special basis states, such as the four outliers on the top left in Fig.~\ref{fig:QMBS_9950_N19_k0_ag2_1.6}. 

A zero momentum basis state $|a(k=0)\rangle$ is linear superposition of all possible translations of the representative state $|a\rangle$. For the observed scarred eigenstate the special basis states are
\begin{align}
|a_1\rangle = |10000000\cdots0\rangle \,, \quad |a_2\rangle = |10100000\cdots0\rangle \,,\nn\\
|a_3\rangle = |10101000\cdots0\rangle \,, \quad |a_4\rangle = |10101010\cdots0\rangle \,,
\end{align}
where $1$ denotes spin-up and $0$ spin-down. The special nature of these states is deeply connected with the magnetic part of the Hamiltonian in Eq.~\eqref{eq:H_Ising} and is explained in detail in Appendix~\ref{app:b}, where also another comparison between the scarred eigenstate and a typical eigenstate around the energy $E-E_0=19.42$ is shown.

\begin{figure}[t]  
\centering
\includegraphics[width=0.45\textwidth]{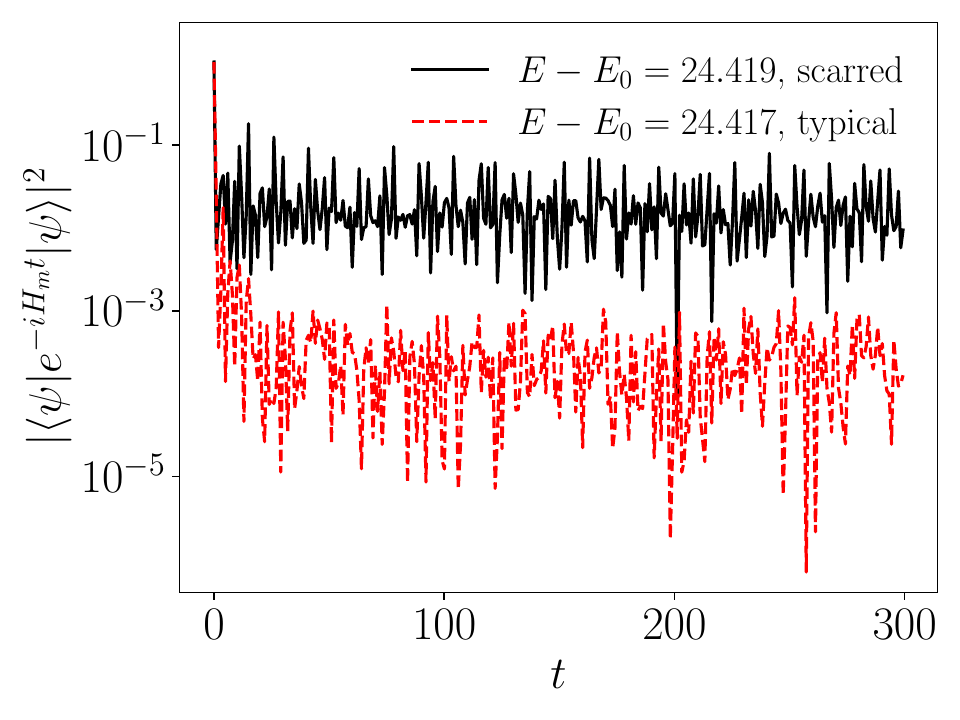}
\caption{Recurrence probabilities of the scarred eigenstate at $E-E_0=24.419$ and a typical eigenstate at $E-E_0=24.417$ under the time evolution driven by the magnetic interaction on a periodic $N=19$ plaquette chain with $j_{\rm max}=\frac{1}{2}$ and $g^2a=1.6$.\footnote{The originally published version has a typo in the figure legend.}}
\label{fig:recurrence}
\end{figure}

Furthermore, we calculate the recurrence probability of an initial scarred eigenstate under the time evolution driven by the magnetic Hamiltonian $|\langle \psi| e^{-iH_mt}| \psi\rangle|^2$ and compare it with that of an initial typical eigenstate of a similar energy. The results are shown in Fig.~\ref{fig:recurrence}, where we see the recurrence probability of the scarred eigenstate at $E-E_0=24.419$ is two orders of magnitude larger than that of the typical eigenstate at $E-E_0=24.417$, which is of the same order as the inverse of the Hilbert space dimension ($27,594$). This difference reflects that the scarred eigenstate does not thermalize since it has large components in those special states that transition into each other preferentially under the magnetic interaction.

\begin{figure}[t]  
\centering
\includegraphics[width=0.45\textwidth]{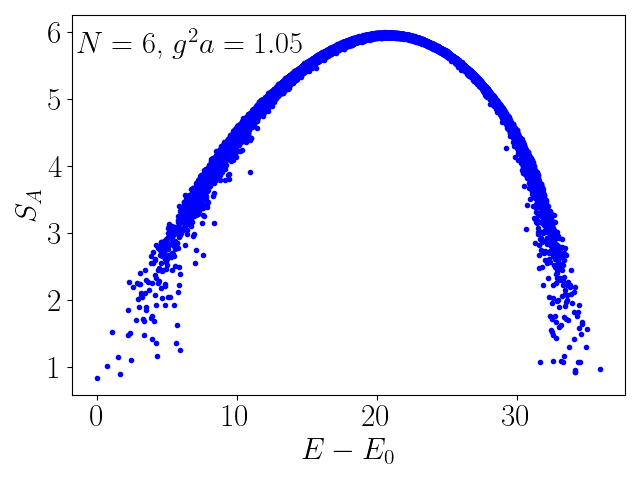}
\caption{Half-chain entanglement entropies for all the energy eigenstates on an aperiodic five-plaquette chain with $j_{\rm max}=1.5$ and $g^2a=1.05$. The closed boundary conditions are $1,1,0,1$ for the external links, which avoid reflection symmetry.}
\label{fig:EE_N5_ag2_1.05_jmax1.5}
\end{figure}

Finally, we discuss whether such quantum many-body scars exist in the continuum limit. Because of the renormalization group flow, the coupling goes to zero in this limit. In order to correctly describe high energy states, one must increase $j_{\rm max}$ simultaneously. In particular, one can show the needed $j_{\rm max}$ value grows as~\cite{Turro:2024pxu}
\begin{align}
\label{eqn:jmax_value}
j_{\rm max} \sim \frac{N_{\rm L} \widetilde{E}}{g^2\epsilon} \,,
\end{align}
where $N_{\rm L}$ denotes the number of links on the lattice, $\widetilde{E}$ is the energy up to which we want to describe all states within an accuracy $\epsilon$ (the definition of $\widetilde{E}$ differs from the energy of the Hamiltonian studied here by an overall shift that makes the magnetic energy non-negative). We study a five-plaquette chain with $j_{\rm max}=1.5$ and $g^2a=1.05$ where we observe no entropy outliers in the middle of the spectrum, as shown in Fig.~\ref{fig:EE_N5_ag2_1.05_jmax1.5}. Similar results are obtained for other $N,g^2a,j_{\rm max}$ and boundary conditions (see Appendix~\ref{app:c}). Therefore, we surmise that the scarred eigenstates observed for $j_{\rm max} = \frac{1}{2}$ do not occur in the full SU(2) gauge theory.

{\it Conclusions.} In this letter we calculated the entanglement entropy in Hamiltonian $(2+1)$-dimensional SU(2) lattice gauge theory on plaquette chains. We defined a link-cutting procedure to partition a chain that renders the reduced density matrix and the entanglement entropy invariant under time-independent gauge transformations. We showed that the entanglement entropies of the ground state and excited states follow Page curves that fit a simple scaling law (\ref{eq:Scross}). We also calculated the half-chain entanglement entropies for all eigenstates and observed the presence of scars in the middle of the spectrum on a periodic plaquette chain with $j_{\rm max}=\frac{1}{2}$. These scars are a result of the existence of special electric basis states that transform into each other by the action of the magnetic Hamiltonian. We found no highly excited states of exceptionally low entanglement entropy in cases with smaller $g^2$ and higher $j_{\rm max}$. This suggests that in the physical limit, there are no quantum many-body scars in the SU(2) gauge theory.

There are many possible avenues of further exploration of the entanglement properties of SU(2) gauge theory on the lattice. Such studies would include the effects of finite volume and electric flux truncation, the time evolution of information scrambling, and the modular (entanglement) Hamiltonian. We plan to pursue these studies in future work.

\begin{acknowledgments}
X.Y. thanks Niklas Mueller for interesting discussions. The authors gratefully acknowledge the scientific support and HPC resources provided by the Erlangen National High Performance Computing Center (NHR@FAU) of the Friedrich-Alexander-Universität Erlangen-Nürnberg (FAU).
B.M. acknowledges support by the U.S. Department of Energy, Office of Science (grant DE-FG02-05ER41367).
X.Y. is supported by the U.S. Department of Energy, Office of Science, Office of Nuclear Physics, InQubator for Quantum Simulation (IQuS) (https://iqus.uw.edu) under Award Number DOE (NP) Award DE-SC0020970 via the program on Quantum Horizons: QIS Research and Innovation for Nuclear Science. 
\end{acknowledgments}

\bibliographystyle{apsrev4-1}
\bibliography{main}

\begin{widetext}
\appendix

\section{Simplification of the entanglement entropy calculation when $j_{\rm max}=\frac{1}{2}$}
\label{app:a}

As mentioned in the main text, when $j_{\rm max}=\frac{1}{2}$, the plaquette chain can be mapped onto an Ising chain, where states can be represented by spins pointing up or down at each plaquette location. In the Ising representation, degrees of freedom live at each spin site and the division into subsystems is natural; one just assigns a certain number of spin sites to be a subsystem and then the calculation of entanglement entropy is straightforward. This has been widely studied. Here we study it because it corresponds to the SU(2) lattice gauge theory in the strong coupling limit, i.e., $j_{\rm max}=\frac{1}{2}$ and the structure of the $\sigma_i^x$ term in Eq.~\eqref{eq:H_Ising} is a result of the SU(2) group property and local Gauss law.

On a periodic spin chain, an eigenstate can be written as
\begin{align}
|\psi(k)\rangle = \sum_a c_a |a(k)\rangle \,,
\end{align}
where the momentum basis state is defined as
\begin{align}
|a(k)\rangle = \frac{1}{\sqrt{N_a}}\sum_{r=0}^{N-1} e^{-ikr} \hat{T}^r |a\rangle \,,
\end{align}
with a representative state $|a\rangle$ of the translation $\hat{T}$ equivalent class. The reduced density matrix of a subchain $A$ is
\begin{align}
\rho_A = {\rm Tr}_{A^c}( |\psi(k)\rangle \langle \psi(k)| ) = \sum_{a,b} c_a^* c_b\, {\rm Tr}_{A^c}( |b(k)\rangle \langle a(k)| ) = \sum_{a,b} \frac{c_a^* c_b}{\sqrt{N_aN_b}} \sum_{r_a,r_b} e^{ik(r_a-r_b)} {\rm Tr}_{A^c}( \hat{T}^{r_b} |b\rangle \langle a| \hat{T}^{-r_a} ) \,.
\end{align}
In the calculation, one can save ${\rm Tr}_{A^c}( \hat{T}^{r_b} |b\rangle \langle a| \hat{T}^{-r_a} )$ for all $r_a,r_b,|a\rangle$ and $|b\rangle$ for repeated use.

The reader may wonder why in the Ising representation there seems to exist no edge effect as in the general case, where the two cut links could not be uniquely assigned to one subsystem. It turns out that when $j_{\rm max}=\frac{1}{2}$, the $j$ values on the vertical links are redundant when we only consider states with no electric flux winding around the periodic chain. (States with winding electric flux are decoupled from those with none). Spin-up (spin-down) at a site means the top and bottom links in the corresponding plaquette are both in $j=\frac{1}{2}$ ($j=0$). As a result of Gauss law the $j$ value on a vertical link is determined by the $j$ values on the two top or bottom links that share the same vertex with the vertical link. When we divide the spin chain into left and right segments, we actually cut two vertices along a vertical link in the original plaquette chain. This is only possible without ambiguity when $j_{\rm max}=\frac{1}{2}$. For example, we consider $j_{\rm max}>\frac{1}{2}$ and a vertex with three links: left, right and vertical. If the right and vertical links have $j=\frac{1}{2}$, the left link can have either $j=1$ or $j=0$. In either case, although the $j$ values on the right and vertical links look identical, they actually correspond to different states: When the left link has $j=1$, the two $j=\frac{1}{2}$ states on the right and vertical links are symmetric, while they are antisymmetric when the left link has $j=0$. We note that if we had kept the $m_L,m_R$ quantum numbers in the basis states, cutting through vertices would be possible without causing an ambiguity. 

\section{More results for $j_{\rm max}=\frac{1}{2}$}
\label{app:b}

\subsection{Non-integrability of the system when $j_{\rm max}=\frac{1}{2}$ and $g^2a=1.6$}

As mentioned in the main text, the periodic $N=17$ plaquette chain with $j_{\rm max}=\frac{1}{2}$ is non-integrable when $g^2a=1.6$. We confirm this by analysing level statistics. It is known that the distribution of eigenenergy gaps in non-integrable systems follows the Wigner-Dyson statistics featuring a vanishing probability of zero gap (level repulsion). In the following level statistics analysis, we will exclude the lowest $1500$ and the highest $1500$ eigenstates (ordered by their eigenenergies) to reduce the spectrum boundary effect.

\begin{figure}[t]  
\centering
\subfloat[Rescaled energy gap.\label{fig:wd_scaled_N17_ag2_1.6}]{%
  \includegraphics[width=0.45\linewidth]{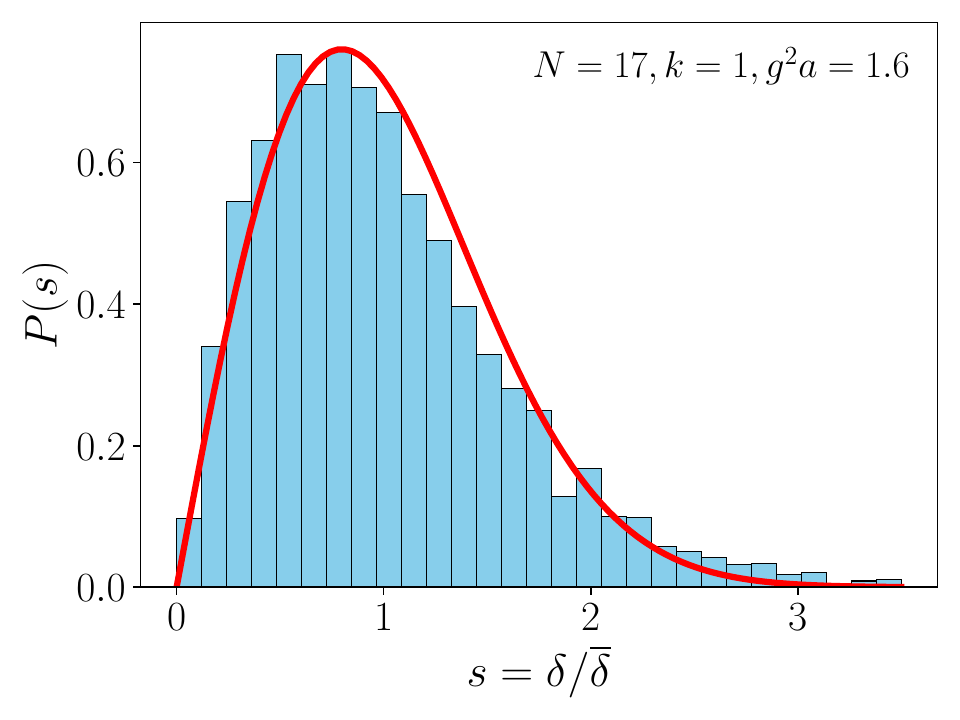}%
}~~~~~~
\subfloat[Restricted gap ratio.\label{fig:gap_ratio_N17_ag2_1.6}]{%
  \includegraphics[width=0.45\linewidth]{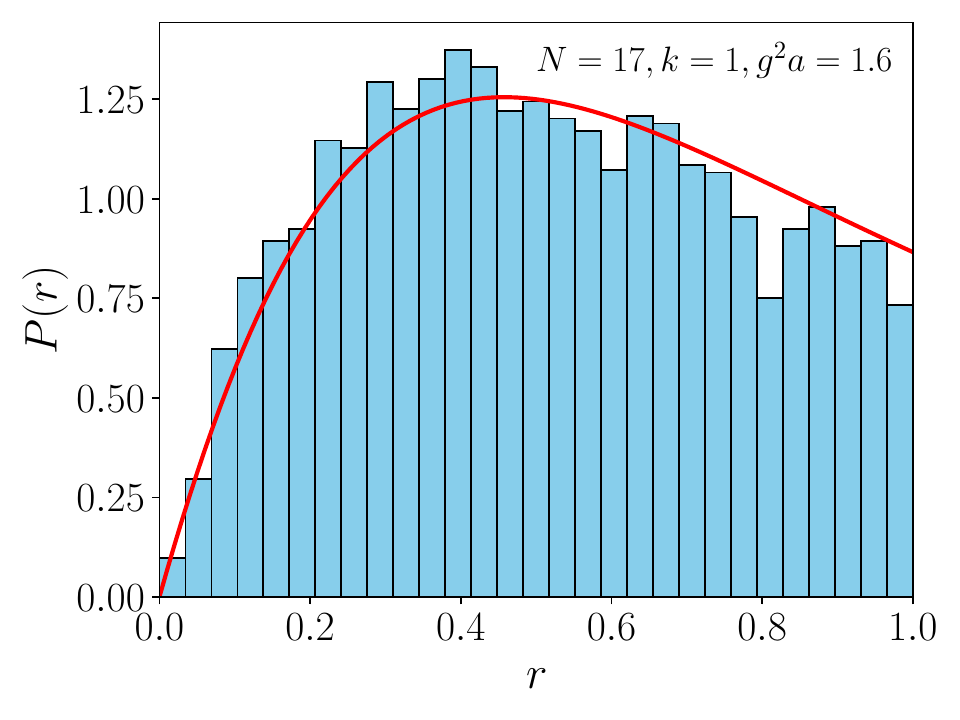}%
}
\caption{Level statistics in the $k=1$ sector on a periodic $N=17$ plaquette chain with $j_{\rm max}=\frac{1}{2}$ and $g^2a=1.6$: (a) rescaled energy gap, (b) restricted gap ratio. The red lines are predictions from the Gaussian orthogonal ensemble.}
\label{fig:gap_ratio_wd_scaled}
\end{figure}

In Fig.~\ref{fig:wd_scaled_N17_ag2_1.6}, the distribution of the rescaled energy gap $s\equiv \delta/\bar{\delta}$ is shown, where $\delta$ denotes the energy gap and $\bar{\delta}$ is the average value of the energy gap. The red line is a prediction from $2\times2$ random matrices in the Gaussian orthogonal ensemble (GOE). We used the $k=1$ momentum sector (the momentum is $\frac{2\pi}{N}k$) since the $k=0$ sector has a discrete parity symmetry that must be lifted in order to expose the repulsive level statistics. Furthermore, the restricted gap ratio defined as
\begin{align}
0 < r_\alpha = \frac{{\rm min}[\delta_\alpha,\delta_{\alpha-1}]}{{\rm max}[\delta_\alpha,\delta_{\alpha-1}]} \leq 1\,,
\end{align}
is depicted in Fig.~\ref{fig:gap_ratio_N17_ag2_1.6}, where the red curve is a prediction from the GOE. The expectation value of the restricted gap ratio is $\langle {r}\rangle\approx0.518$, which is very close to the GOE prediction $\langle r \rangle_{\rm GOE} \approx 0.5307$. Based on these analyses, we conclude the system is non-integrable when $j_{\rm max}=\frac{1}{2}$ and $g^2a=1.6$.

\subsection{Relation between electric and magnetic energies when $j_{\rm max}=\frac{1}{2}$ and $g^2a=1.6$}

In the main text, $g^2a=1.6$ is used when studying quantum many-body scars in the case of $j_{\rm max}=\frac{1}{2}$. We choose this value since at it the equal partition of the electric and magnetic energies is approximately valid, which is a property of the ($2+1$)-dimensional SU(2) gauge theory at thermal equilibrium in the continuum. The relation between the electric and magnetic energies for all the eigenstates is shown in Fig.~\ref{fig:He_vs_Hm_N17_k0_ag2_1.6}, where we have subtracted the corresponding energies of the ground state, which makes the approximate equality manifest. A few isolated and mixed bands seen in the plot are a remnant of the isolated electric energy sectors which will be discussed later. 
Some degree of spreading is still seen in the plot, which is caused by the local Hilbert space truncation effects. As will be shown later, increasing $j_{\rm max}$ shrinks the spreading.

\begin{figure}[h]  
\centering
\includegraphics[width=0.45\textwidth]{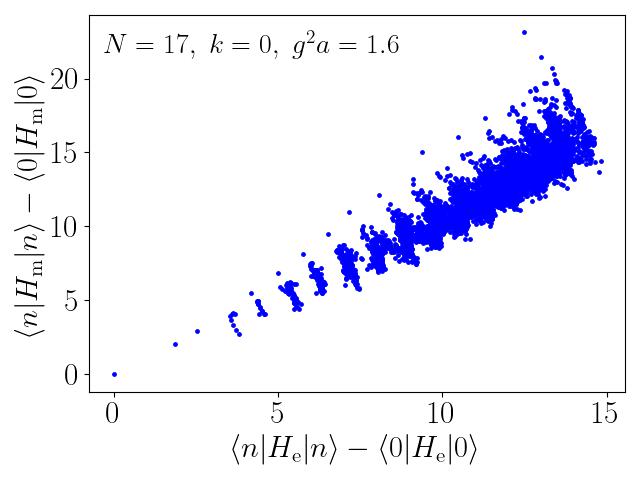}
\caption{Relation between the electric and magnetic energies of all the eigenstates relative to those of the ground state in the $k=0$ sector on a periodic $N=17$ plaquette chain with $j_{\rm max}=\frac{1}{2}$ and $g^2a=1.6$.}
\label{fig:He_vs_Hm_N17_k0_ag2_1.6}
\end{figure}

\subsection{Scarred eigenstates and special basis states}
In the main text, we show that the scarred eigenstate has large components in some special basis states. Here we show that the transition between these states through the magnetic part of the Hamiltonian dominates over other transitions.  
The magnetic part of the Hamiltonian can be written as~\cite{Yao:2023pht}
\begin{align}
H_m = -2h_x \sum_{i=0}^{N-1} \frac{1-3\sigma_{i-1}^z}{4} \sigma_i^x \frac{1-3\sigma_{i+1}^z}{4} \,.
\end{align}
We consider a tower of states generated by applying $H_m$ ($H_e$ does not change the state) repeatedly to the bare vacuum state $|a_0\rangle = |00000000\cdots0\rangle$. Applying once we get $|a_1(k=0)\rangle$, which is defined in the main text. Due to the $\frac{1-3\sigma_{i\pm1}^z}{4}$ factor, the transition amplitude from $|a_1(k=0)\rangle$ to the state $|11000000\cdots0(k=0)\rangle$ is suppressed by a factor of two compared to that to other states. If we consider the state $|a_2(k=0)\rangle$ among all the states generated by $H_m|a_1(k=0)\rangle$, its transition amplitude to the state $|a_3(k=0)\rangle$ is enhanced at least by a factor of two due to the symmetry compared to the other cases. Whether the ``$10$'' is generated on the left or right next to the ``$1010$'' part in $|a_2(k=0)\rangle$, it leads to the same state due to the translation, which is not the case for other states. For example, $|10100010\cdots0(k=0)\rangle$ and $|10001010\cdots0(k=0)\rangle$ are two different states, but $|10101000\cdots0(k=0)\rangle$ and $|101000\cdots0010(k=0)\rangle$ correspond to the same state. [If one lifts up the parity symmetry and focus on the parity-even sector, the transition from $|a_2(k=0)\rangle$ to $|a_3(k=0)\rangle$ is still enhanced by a factor of $\sqrt{2}$ compared to that to the state $(|10100010\cdots0(k=0)\rangle + |10001010\cdots0(k=0)\rangle)/\sqrt{2}$.] Again the transition amplitude from $|a_3(k=0)\rangle$ to $|a_4(k=0)\rangle$ is enhanced by at least a factor of two. This pattern continues until the string ``$101010\cdots$'' exceeds half of the lattice size, when the total length of the lattice and the oddness of the total lattice site number start to matter. The transitions among these $|a_i(k=0)\rangle$ states dominate in the semiclassical limit, where they form a periodic orbit whose length grows linearly with the lattice size. If an eigenstate has a large overlap with these special states, which is the case in Fig.~\ref{fig:QMBS_9950_N19_k0_ag2_1.6}, it is expected that its entanglement entropy will only grow logarithmically with the subsystem size, since the dimension of the space spanned by these special states is just linear in $N$ rather than exponential in $N$.

In Fig.~\ref{fig:scar_state2} we compare the wavefunction components of the scarred eigenstate at the energy $E-E_0=19.414$ with those of a typical eigenstate at the energy $E-E_0=19.419$ for a periodic $N=19$ plaquette chain with $j_{\rm max}=\frac{1}{2}$ and $g^2a=1.6$. We find that the scarred eigenstate has large components in the special basis states discussed above and in the main text.

\begin{figure}[t]  
\centering
\subfloat[Scarred eigenstate.\label{fig:QMBS_2436_N19_k0_ag2_1.6}]{%
  \includegraphics[width=0.45\textwidth]{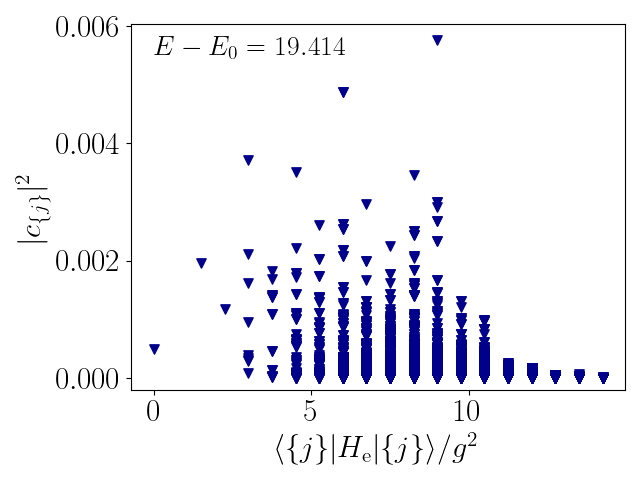}%
}\hfill
\subfloat[Typical eigenstate.\label{fig:QMBS_2440_N19_k0_ag2_1.6}]{%
  \includegraphics[width=0.45\textwidth]{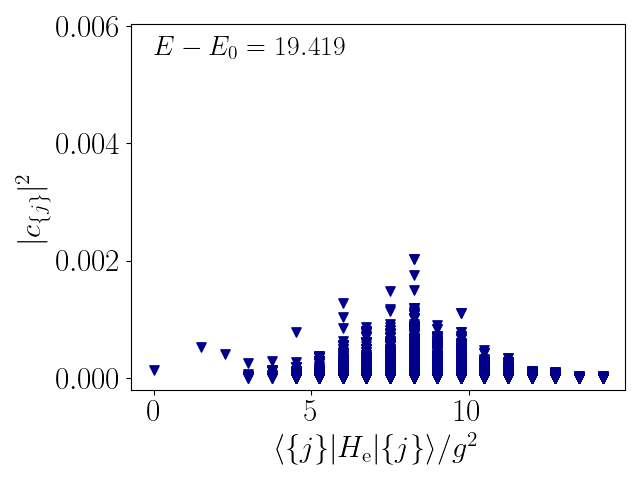}%
}
\caption{Wavefunction components of the scarred eigenstate and a typical eigenstate around the energy $E-E_0=19.42$ in the electric basis in the $k=0$ sector on a periodic $N=19$ plaquette chain with $j_{\rm max}=\frac{1}{2}$ and $g^2a=1.6$. Their half-chain entanglement entropies are $2.70$ and $5.09$, respectively. The $x$-axis labels the electric energy of the basis state which is zero when all $j$ values are $0$. We fix the $y$-axis scale to be the same in both plots for visual comparison.}
\label{fig:scar_state2}
\end{figure}

\subsection{Results at extreme couplings when $j_{\rm max}=\frac{1}{2}$}

\begin{figure}[t]  
\subfloat[\label{fig:gap_ratio_N17_ag2_2.4}]{%
  \includegraphics[width=0.46\linewidth]{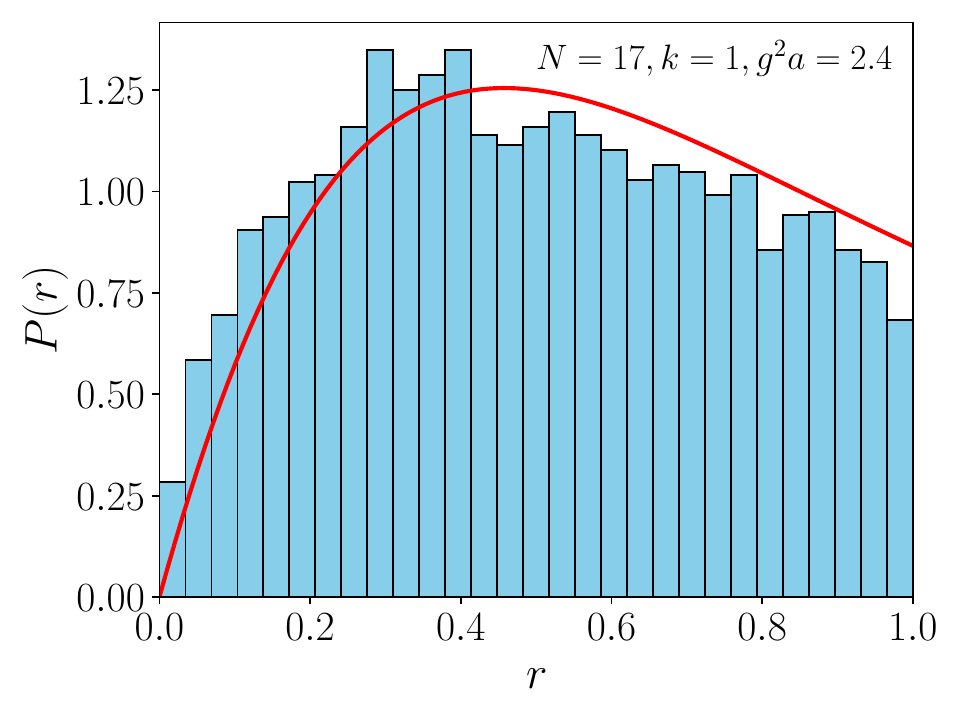}%
}\hfill
\subfloat[\label{fig:gap_ratio_N17_ag2_0.1}]{%
  \includegraphics[width=0.46\linewidth]{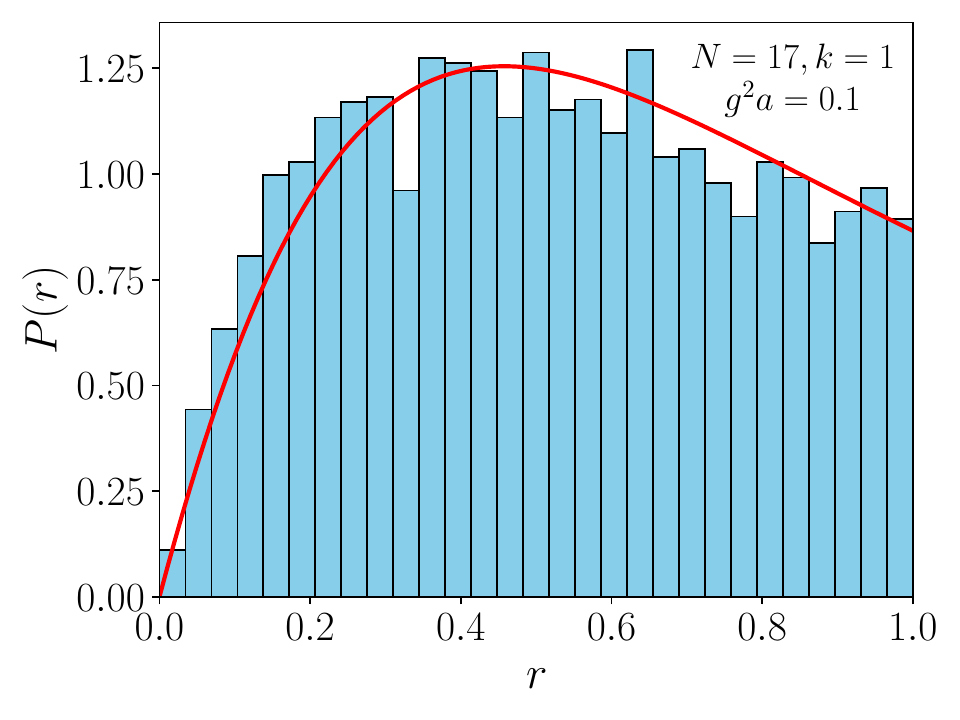}%
}

\subfloat[\label{fig:EE_N17_k0_Nsub8_ag2_2.4}]{%
  \includegraphics[width=0.45\linewidth]{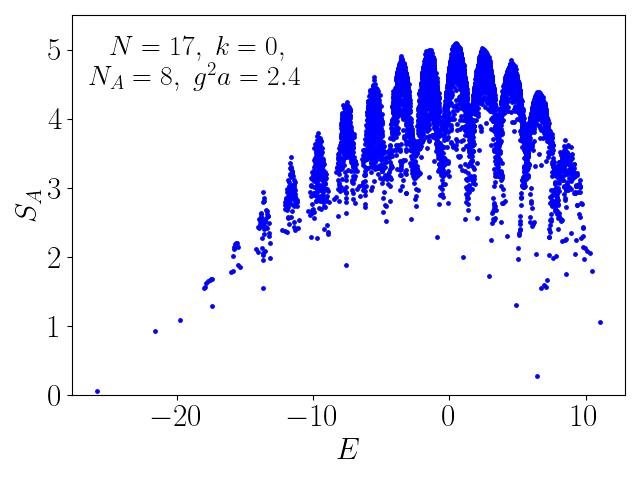}%
}\hfill
\subfloat[\label{fig:EE_N17_k0_Nsub8_ag2_0.1}]{%
  \includegraphics[width=0.45\linewidth]{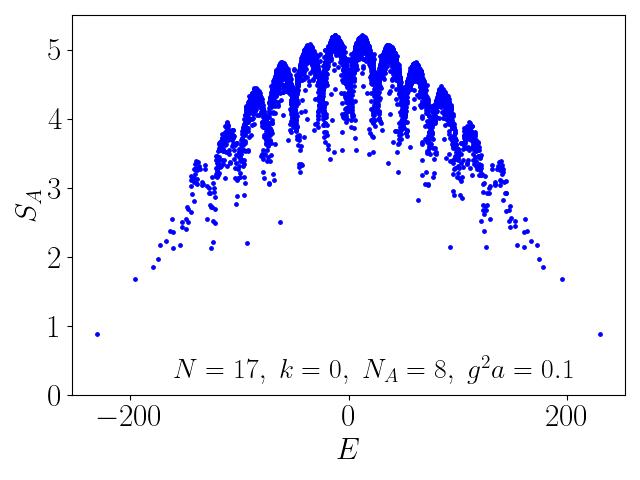}%
}

\subfloat[\label{fig:He_vs_Hm_N17_k0_ag2_2.4}]{%
  \includegraphics[width=0.46\linewidth]{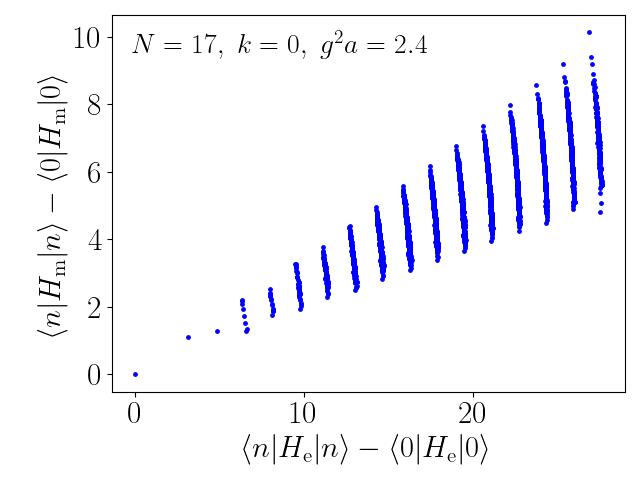}%
}\hfill
\subfloat[\label{fig:He_vs_Hm_N17_k0_ag2_0.1}]{%
  \includegraphics[width=0.46\linewidth]{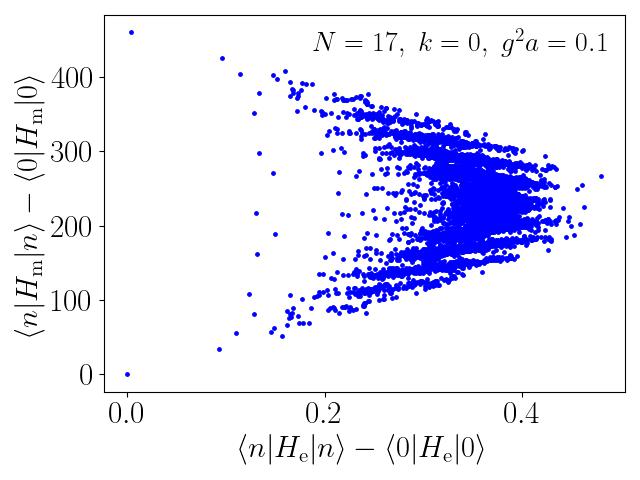}%
}
\caption{Distributions of restricted gap ratios (top), half-chain entanglement entropies (middle) and relations between the electric and magnetic energies of all eigenstates (bottom) on a periodic $17$-chain with $j_{\rm max}=\frac{1}{2}$ at strong ($g^2a=2.4$, left) and weak ($g^2a=0.1$, right) couplings.}
\label{fig:EE_extreme_N17_ag2}
\end{figure}

Here we study the system's behavior at extreme couplings. First we consider the case with a strong coupling constant $g^2a=2.4$ and calculate the distribution of the restricted gap ratio, the half-chain entanglement entropy and the relation between the electric and magnetic energies. The results are shown in Fig.~\ref{fig:EE_extreme_N17_ag2}. In the level statistics analysis, the lowest and highest 1500 states are excluded, as explained above. The feature of level repulsion and the reasonable agreement with the GOE prediction in Fig.~\ref{fig:gap_ratio_N17_ag2_2.4} show the system at this coupling is still non-integrable. But at the same time, we observe a manifest structure of multiple vertical bands in the relation between the electric and magnetic energies, as shown in Fig.~\ref{fig:He_vs_Hm_N17_k0_ag2_2.4}. These bands correspond to different electric energy sectors, in each of which various states have roughly equal electric energy and thus the electric energy corresponds to an approximate symmetry. These electric energy sectors are equally and widely separated in the strong coupling limit. It is then expected that the lowest energy state in each sector only significantly overlaps with states in the same sector and is almost decoupled from other sectors. In other words, the lowest energy state in each sector does not explore the whole Hilbert space when represented in the electric basis $|\{j\}\rangle$. Therefore, the entanglement entropies of these states follow sub-volume laws, shown as the outsiders in Fig.~\ref{fig:EE_N17_k0_Nsub8_ag2_2.4}. Since they have equally separated electric energies, which dominate the total energy in the strong coupling limit, they appear equidistant in (total) energy in the plot. The multiple-arc structure seen in Fig.~\ref{fig:EE_N17_k0_Nsub8_ag2_2.4} corresponds to the decoupled electric energy sectors depicted in Fig.~\ref{fig:He_vs_Hm_N17_k0_ag2_2.4}. To further demonstrate this, the wavefunction components of the scarred eigenstate at $E-E_0=18.3$ are shown in Fig.~\ref{fig:scar_state3}, where the components of a typical eigenstate nearby are also depicted for comparison.

\begin{figure}[t]  
\centering
\subfloat[Scarred eigenstate.\label{fig:QMBS_518_N17_k0_ag2_2.4}]{%
  \includegraphics[width=0.45\textwidth]{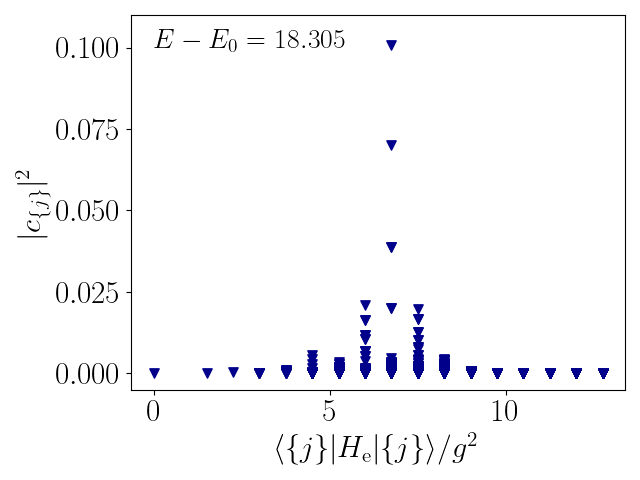}%
}\hfill
\subfloat[Typical eigenstate.\label{fig:QMBS_525_N17_k0_ag2_2.4}]{%
  \includegraphics[width=0.45\textwidth]{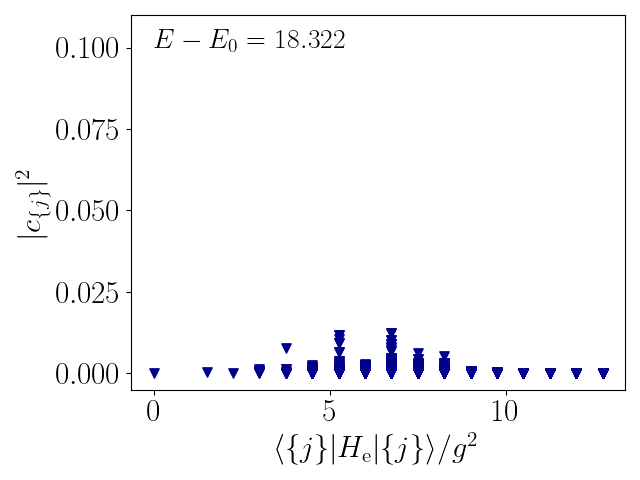}%
}
\caption{Wavefunction components of the scarred eigenstate and a typical eigenstate around the energy $E-E_0=18.3$ in the electric basis in the $k=0$ sector on a periodic $N=17$ plaquette chain with $j_{\rm max}=\frac{1}{2}$ and $g^2a=2.4$. Their half-chain entanglement entropies are $1.89$ and $3.62$, respectively. The $x$-axis labels the electric energy of the basis state which is zero when all $j$ values are $0$. We fix the $y$-axis scale to be the same in both plots for visual comparison.}
\label{fig:scar_state3}
\end{figure}

As a comparison, we also show results in the weak coupling limit in Fig.~\ref{fig:EE_extreme_N17_ag2}. We choose $g^2a=0.1$ at which the system is still non-integrable, as shown in Fig.~\ref{fig:gap_ratio_N17_ag2_0.1}. We would expect equally and widely separated magnetic energy sectors to appear which correspond to a multiple-arc structure in the entanglement entropy plot. The multiple-arc structure can be seen in Fig.~\ref{fig:EE_N17_k0_Nsub8_ag2_0.1}. However, since we use the electric basis and truncate the basis at $j_{\rm max}=\frac{1}{2}$, the decoupled magnetic sectors are not very pronounced in Fig.~\ref{fig:He_vs_Hm_N17_k0_ag2_0.1}. This also makes the low-entanglement states in Fig.~\ref{fig:EE_N17_k0_Nsub8_ag2_0.1} not as manifest as in the strong coupling case shown in Fig.~\ref{fig:EE_N17_k0_Nsub8_ag2_2.4}.

\section{More results with $j_{\rm max}>\frac{1}{2}$}
\label{app:c}

\begin{figure*}[p]  
\centering
\subfloat[\label{fig:EE_N5_ag2_1.2_jmax1}]{%
  \includegraphics[width=0.33\linewidth]{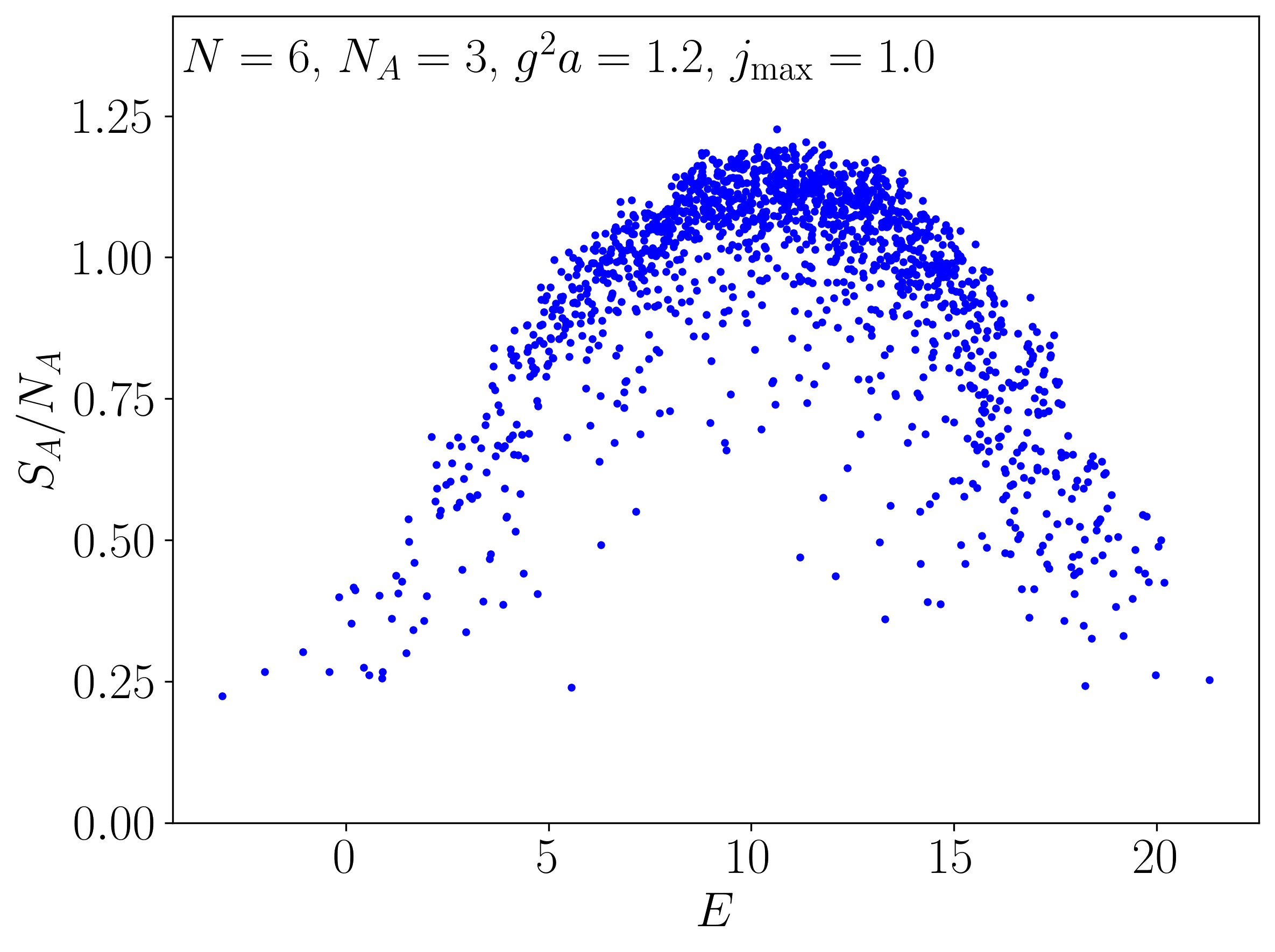}%
}\hfill
\subfloat[\label{fig:EE_N7_ag2_1.2_jmax1}]{%
  \includegraphics[width=0.33\linewidth]{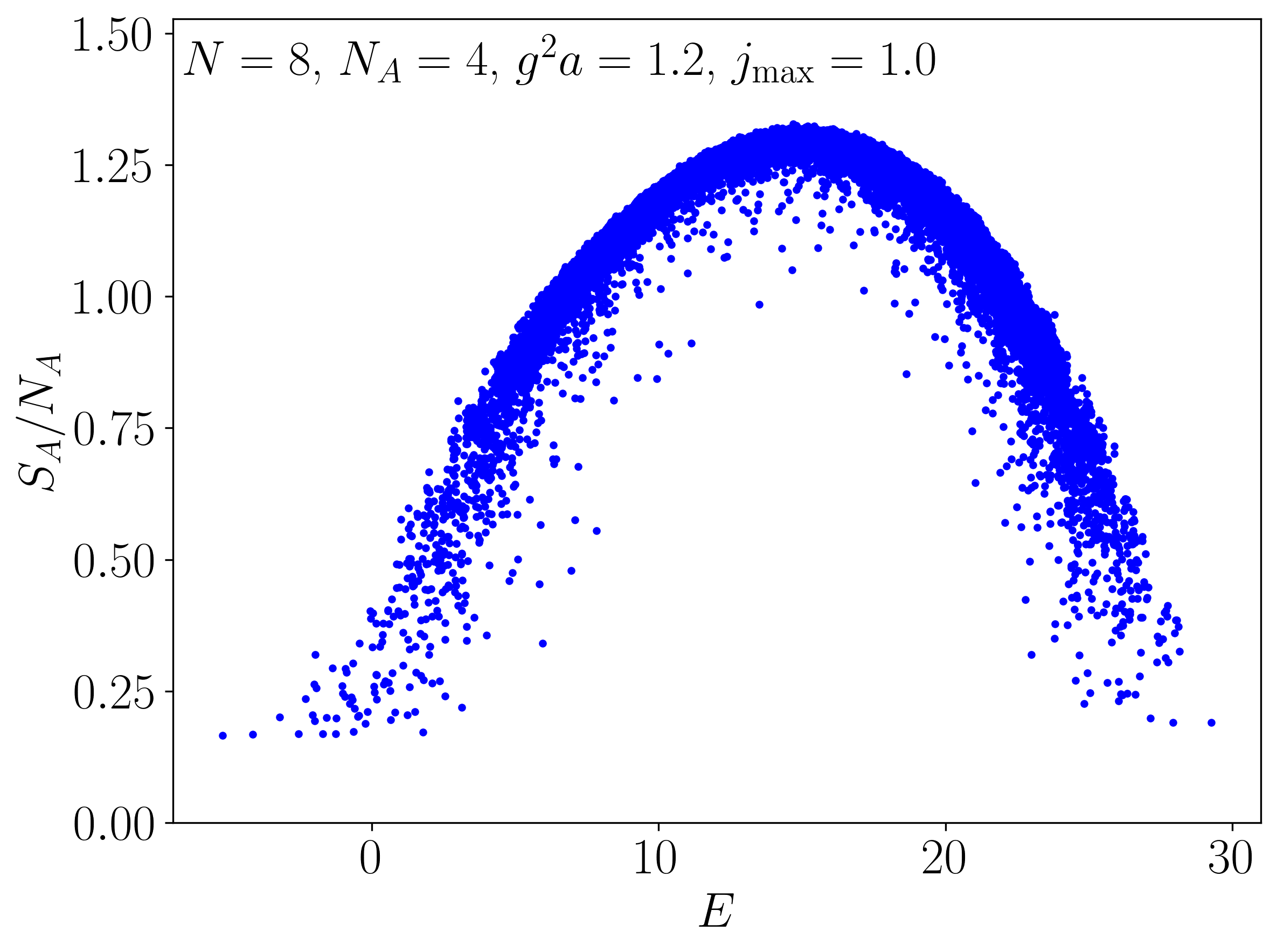}%
}\hfill
\subfloat[\label{fig:EE_N5_ag2_1.0_jmax1.5}]{%
  \includegraphics[width=0.33\linewidth]{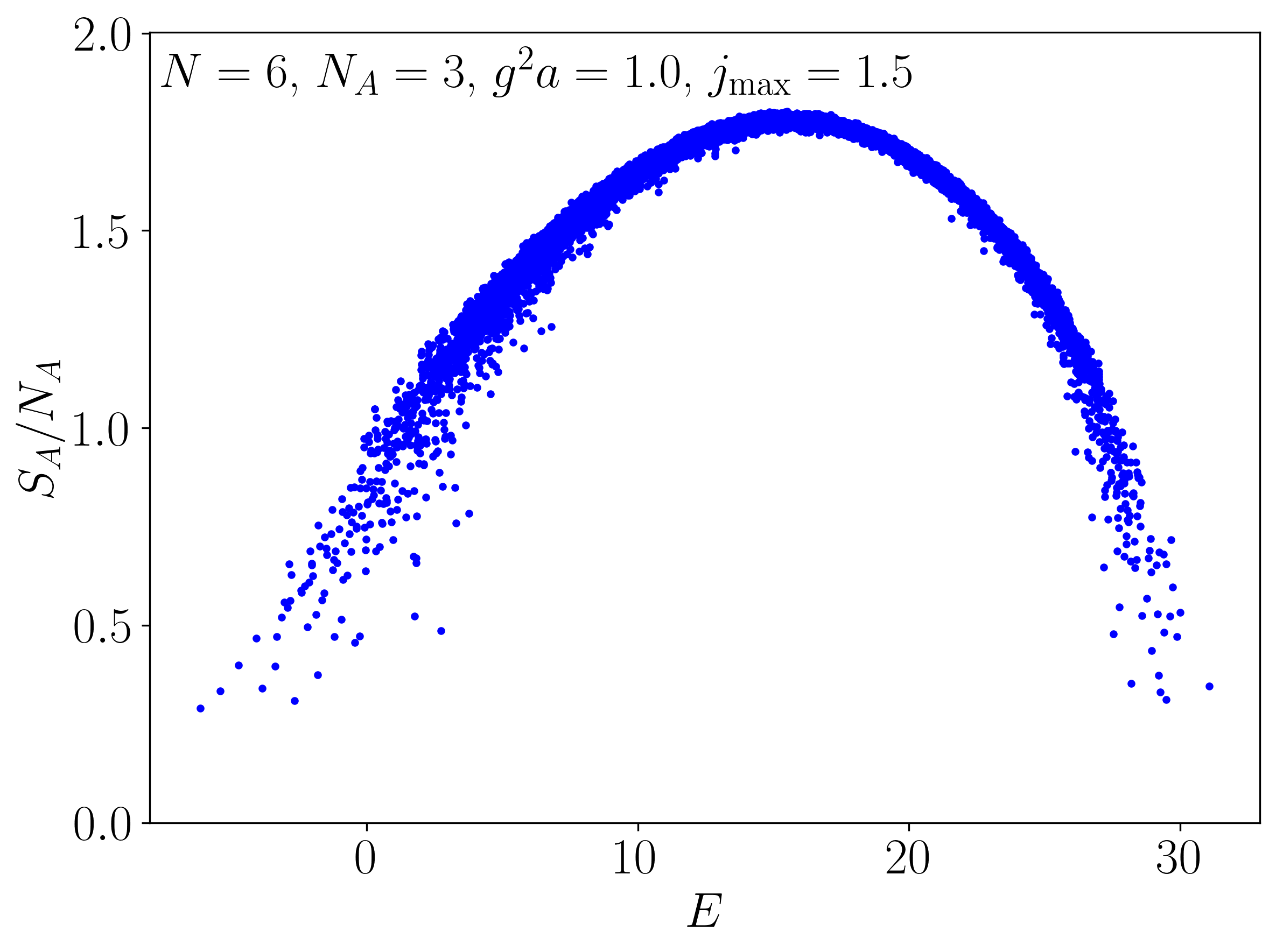}%
}

\centering
\subfloat[\label{fig:He_vs_Hm_N5_ag2_1.2_jmax1}]{%
  \includegraphics[width=0.33\linewidth]{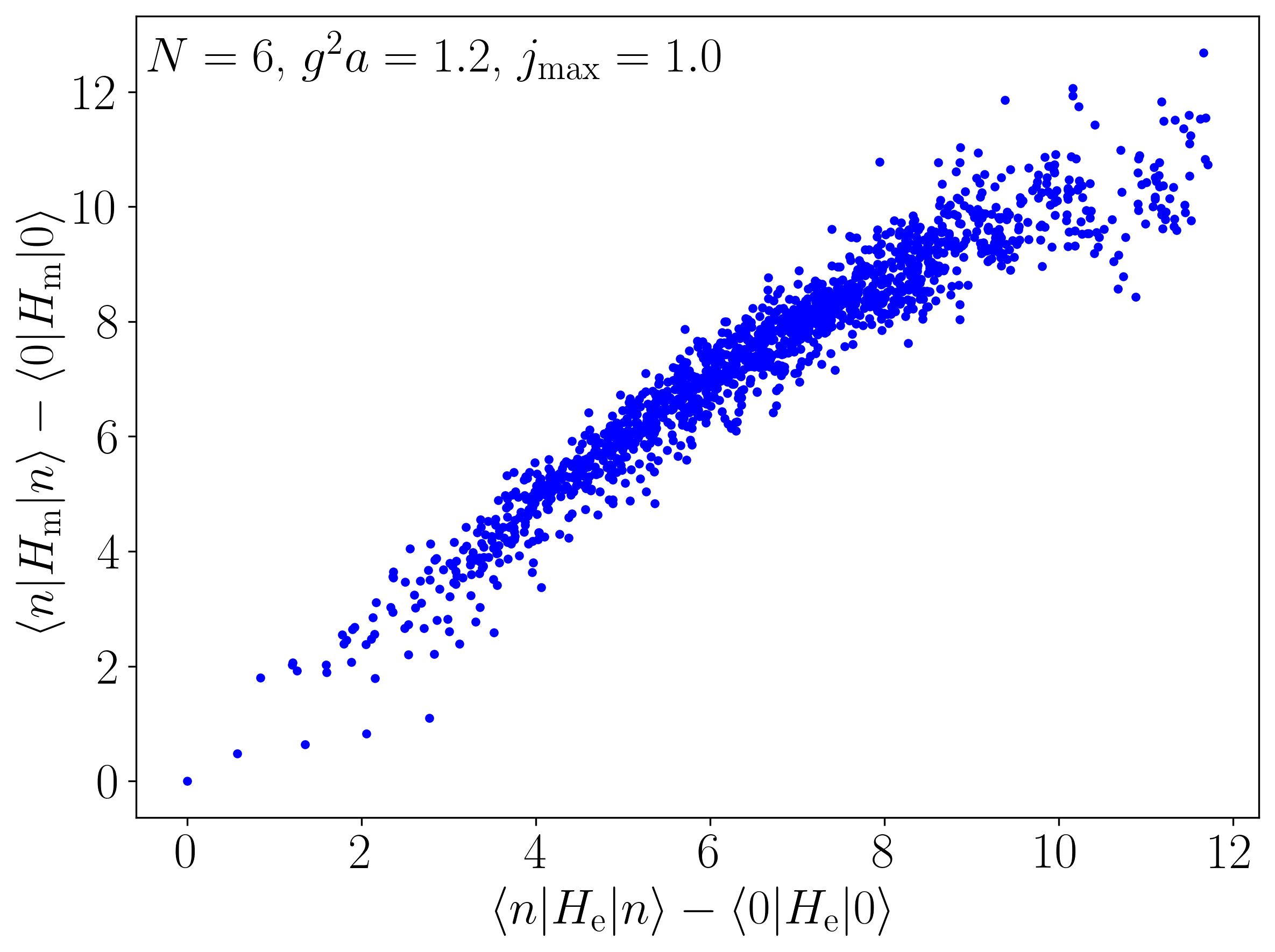}%
}\hfill
\subfloat[\label{fig:He_vs_Hm_N7_ag2_1.2_jmax_1}]{%
  \includegraphics[width=0.33\linewidth]{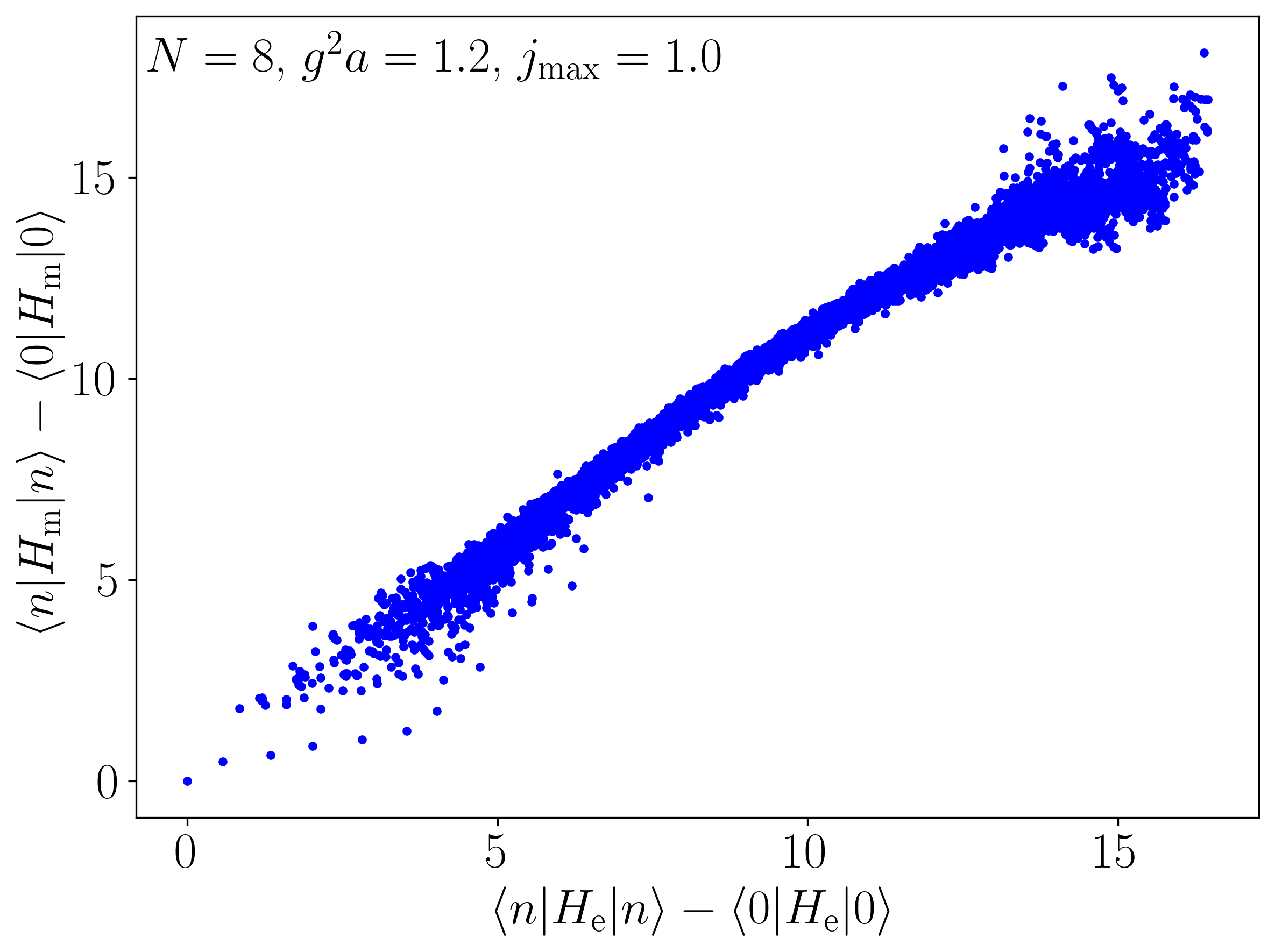}%
}\hfill
\subfloat[\label{fig:He_vs_Hm_N5_ag2_1.0_jmax_1.5}]{%
  \includegraphics[width=0.33\linewidth]{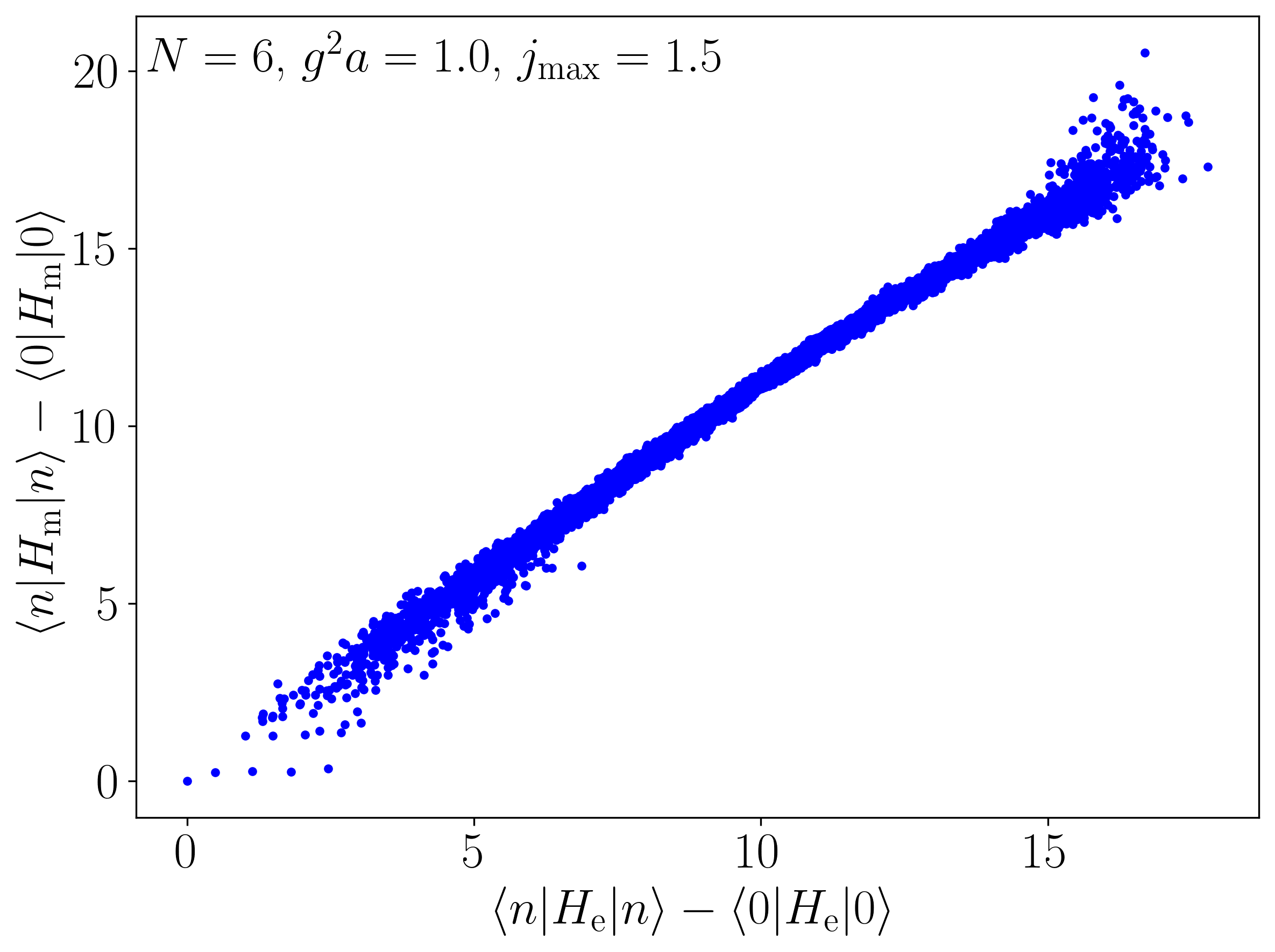}%
}
\caption{Half-chain entanglement entropies (top) and relations between the electric and magnetic energies of all eigenstates (bottom) on a five-plaquette chain with $j_{\rm max}=1$ (left), $j_{\rm max}=1.5$ (right) and on a seven-plaquette chain with $j_{\rm max}=1$ (middle).
The following closed boundary condition is used for the four external links: $j_1=1, j_2=0, j_3=0, j_4=0$.}
\label{fig:EE_He_vs_Hm_N5/7_jmax1/1.5}
\end{figure*}

The existence of special states for the $j_{\rm max}=\frac{1}{2}$ case with unusual high transition amplitudes among each other seems to break down for $j_{\rm max} > \frac{1}{2}$. The plots in Fig.~\ref{fig:EE_He_vs_Hm_N5/7_jmax1/1.5} further support this observation, as some characteristic scar states do not disappear when the length of the chain is increased from five to seven plaquettes, but they disappear completely when the local Hilbert space basis is expanded to $j_{\rm max}=\frac{3}{2}$.

\begin{figure}  
\centering
\subfloat[\label{fig:EE_N6_ag2_1_jmax1.5_parity}]{%
  \includegraphics[width=0.45\linewidth]{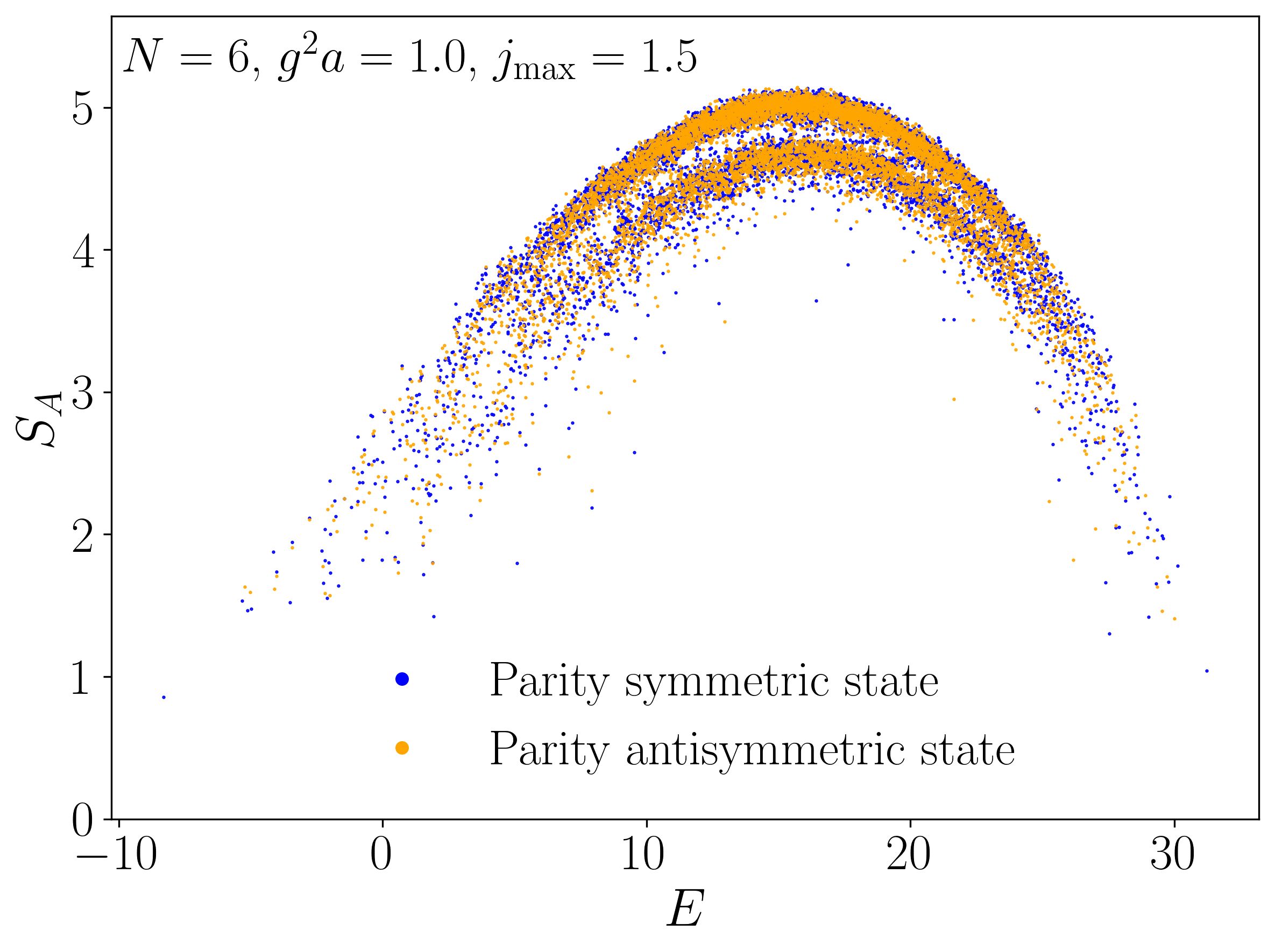}%
}
\subfloat[\label{fig:EE_N6_ag2_1_jmax1.5_TB}]{%
  \includegraphics[width=0.45\linewidth]{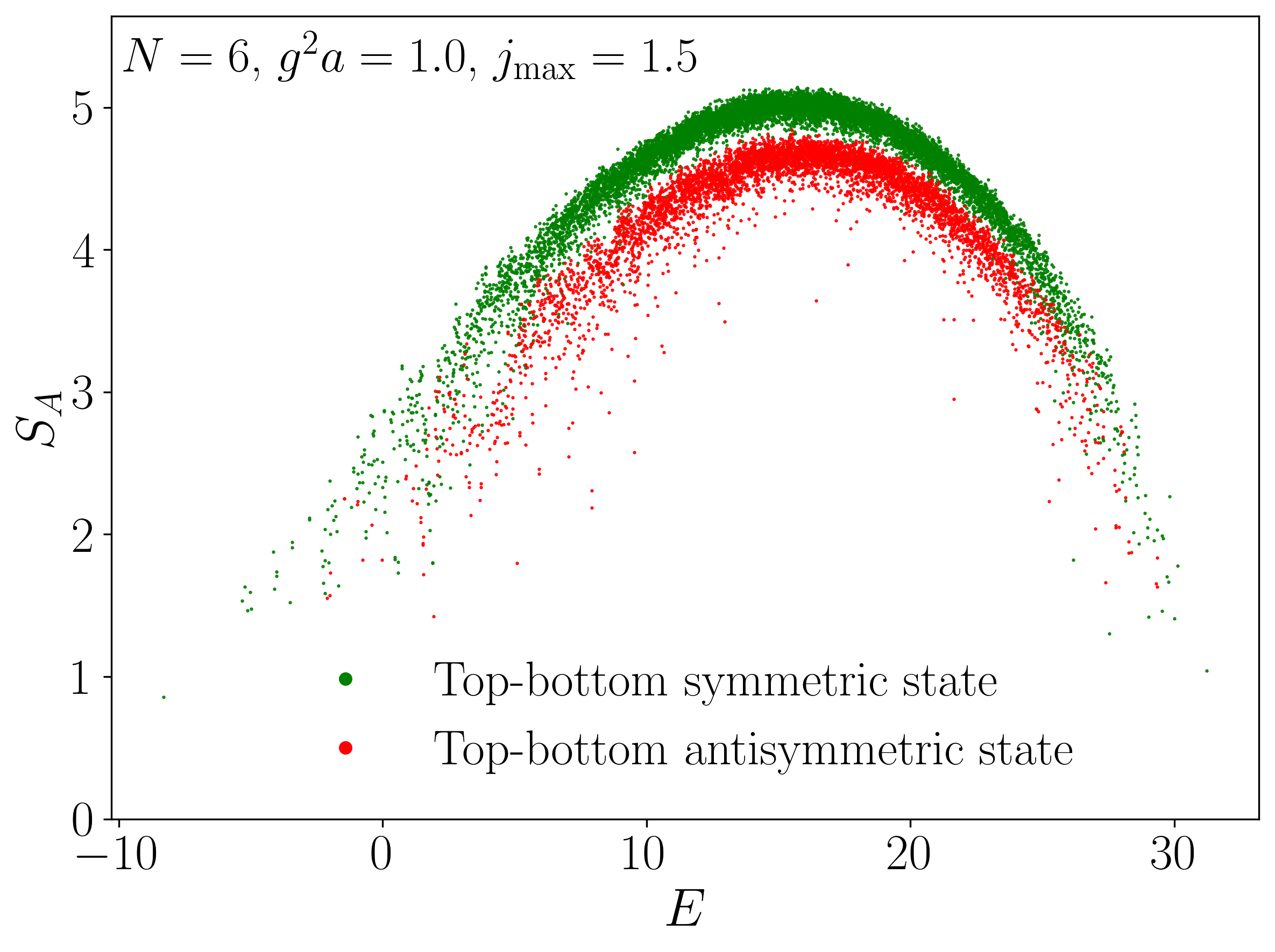}%
}

\caption{Half-chain entanglement entropies of all eigenstates on a five-plaquette chain with $j_{\rm max}=1.5$ at $g^2a=1.0$ and closed boundary conditions $j_1=j_2=j_3=j_4=0$, which preserve the left-right and top-bottom symmetries. The eigenvalues are marked in color depending on whether the corresponding eigenvector is symmetric or antisymmetric with respect to (a) the parity (left-right) transformation or (b) the top-bottom transformation.}
\label{fig:EE_He_vs_Hm_N5_ag2_1_jmax1.5_0_0_0_0}
\end{figure}

So far, we only considered boundary conditions that break parity and top-bottom reflection symmetries. In Fig.~\ref{fig:EE_He_vs_Hm_N5_ag2_1_jmax1.5_0_0_0_0} we show results when both symmetries are restored. One main difference from the other cases is that two dominant arcs show up in the entanglement entropy plot. 

The explanation of the two-arc structure is connected with the symmetry properties of the eigenstate corresponding to the respective eigenenergy. In Fig.~\ref{fig:EE_N6_ag2_1_jmax1.5_parity} the parity symmetric states are colored blue and the antisymmetric states are colored orange. No preference for one of the symmetries can be recognized in this plot for either of the two arcs. In Fig.~\ref{fig:EE_N6_ag2_1_jmax1.5_TB} the top-bottom symmetric states are colored green and the antisymmetric states are colored red. Now a clear difference in the symmetry properties for eigenstates in the two arcs is apparent. We conclude that the top-bottom symmetry is the separating feature of the two arcs. It is interesting that only the top-bottom symmetry leaves an imprint on the entanglement entropy. This may be related to the fact that the boundary of the subsystem coincides with the parity symmetry axis of the plaquette chain.

Another difference of the case with $j_{\rm ext}=0,0,0,0$ external links from cases with other boundary conditions, like the case where the four external links have $1,0,0,0$, is that some scar states seem to appear as shown in Fig.~\ref{fig:EE_He_vs_Hm_N5_ag2_1_jmax1.5_0_0_0_0}.
To see them disappear, one needs to use even higher $\jmax$.
This expectation is supported by the fact that the scars disappear from the middle of the spectrum when $\jmax$ is increased to $2.5$ on a five-plaquette chain.

In Fig.~\ref{fig:He_vs_Hm_N5_ag2_1.05_jmax1.5} we also show the relation between the electric and magnetic energies for the eigenstates on the five-plaquette chain with $\jmax=1.5$, $g^2a=1.05$ and boundary conditions $1,1,0,1$ presented in the main text. As stated above, we find the mixed bands disappear and the spreading decreases.

\begin{figure}[h] 
\centering
\includegraphics[width=0.45\textwidth]{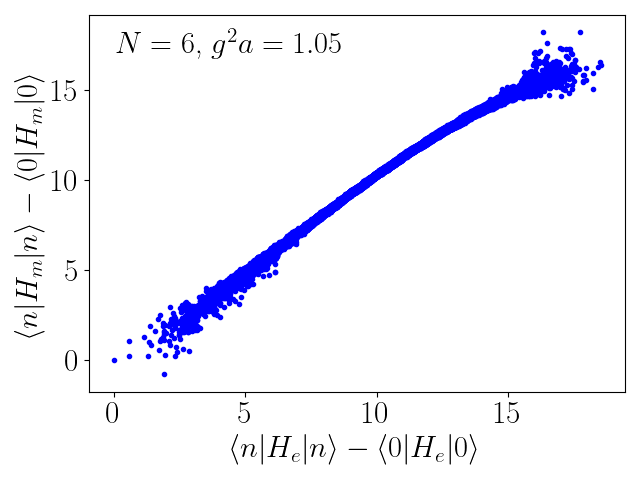}%
\caption{Relation between the electric and magnetic energies for all the energy eigenstates on an aperiodic five-plaquette chain with $j_{\rm max}=1.5$ and $g^2a=1.05$. The closed boundary conditions are $1,1,0,1$ for the external links, which avoid left-right and top-bottom symmetries.}
\label{fig:He_vs_Hm_N5_ag2_1.05_jmax1.5}
\end{figure}

\end{widetext}

\vfill

\end{document}